\documentclass[%
 reprint,
superscriptaddress,
 amsmath,amssymb,
 aps,
]{revtex4-1}

\usepackage{graphicx}
\usepackage{dcolumn}
\usepackage{bm}
\usepackage{siunitx}

\begin{document}

\preprint{}

\title{Estimation of spin-orbit torques in the presence of current-induced magnon creation and annihilation}%

\author{Paul Noel}
\email{paul.noel@ipcms.unistra.fr\\
Present address: Université de Strasbourg, CNRS, Institut de Physique et Chimie des Matériaux de Strasbourg, UMR 7504, Strasbourg F-67000, France}
\affiliation{Department of Materials, ETH Zurich, CH-8093 Zurich, Switzerland}
\author{Emir Karadža}
\affiliation{Department of Materials, ETH Zurich, CH-8093 Zurich, Switzerland}
\author{Richard Schlitz}
\affiliation{Department of Materials, ETH Zurich, CH-8093 Zurich, Switzerland}
\author{Pol Welter}
\affiliation{Department of Physics, ETH Zurich, CH-8093 Zurich, Switzerland}
\author{Charles-Henri Lambert}
\affiliation{Department of Materials, ETH Zurich, CH-8093 Zurich, Switzerland}
\author{Luca Nessi}
\affiliation{Department of Materials, ETH Zurich, CH-8093 Zurich, Switzerland}
\affiliation{Dipartimento di Fisica, Politecnico di Milano, Via G. Colombo 81, 20133 Milano, Italy}
\author{Federico Binda}
\affiliation{Department of Materials, ETH Zurich, CH-8093 Zurich, Switzerland}
\author{Christian L. Degen}
\affiliation{Department of Physics, ETH Zurich, CH-8093 Zurich, Switzerland}
\author{Pietro Gambardella}
\email{pietro.gambardella@mat.ethz.ch}
\affiliation{Department of Materials, ETH Zurich, CH-8093 Zurich, Switzerland}

\date{\today}

\begin{abstract}

We present a comprehensive set of harmonic resistance measurements of the dampinglike (DL) and fieldlike (FL) torques in Pt/CoFeB, Pt/Co, W/CoFeB, W/Co, and YIG/Pt bilayers complemented by measurements of the DL torque using the magneto-optical Kerr effect and calibrated by nitrogen vacancy magnetometry on the same devices. The magnon creation-annihilation magnetoresistances depend strongly on temperature and on the magnetic and transport properties of each bilayer, affecting the estimate of both the DL and FL torque. The DL torque, the most important parameter for applications, is overestimated by a factor of 2 in W/CoFeB and by one order of magnitude in YIG/Pt when not accounting for the magnonic contribution to the planar Hall resistance. We further show that the magnonic contribution can be quantified by combining measurements of the nonlinear longitudinal and transverse magnetoresistances, thus providing a reliable method to measure the spin-orbit torques in different material systems.

\end{abstract}

\maketitle

Materials with high spin-orbit torque (SOT) efficiency are essential for the development of new spin-orbitronic devices that enable storage and logic technologies that are fast, nonvolatile, and endurant~\cite{manchon19, grimaldi20, shao21,krizakova22}. The proper metrology of these SOTs is pivotal for the technological development and for the understanding of their underlying physical mechanisms. Several techniques have been developed to quantify the SOTs~\cite{manchon19}, but there appear to be recurrent inconsistencies in the reported results on the SOT efficiency suggesting that the understanding of the involved processes is incomplete, even for the most studied Pt-based systems~\cite{zhu21}. \\

Harmonic Hall resistance measurements are widely used to evaluate SOT effects, and more particularly the efficiency of the dampinglike (DL) torque and the fieldlike (FL) torque in normal metal (NM)/ferromagnet (FM) bilayers~\cite{pi10, garello13, kim13, avci14, hayashi14}. The usual harmonic Hall resistance analysis assumes a constant magnetization, unaffected by the current and magnetic field. However, this reasoning does not hold in general, as the creation or annihilation of magnons induced by the spin current affects the magnetization~~\cite{demidov11, demidov17, borisenko18, thiery18}. As we show in a joint paper~\cite{jointpaper}, the change of magnon population due to spin currents in NM/FM bilayers can give rise to transverse and longitudinal nonlinear magnetoresistances due to magnon creation or annihilation. All magnetoresistances that depend on the magnitude of the magnetization, such as the anisotropic magnetoresistance (AMR), spin Hall magnetoresistance (SMR),  magnon magnetoresistance (MMR), planar Hall effect (PHE), and anomalous Hall effect (AHE), thus have a nonlinear contribution that depends on the current density. We call this set of nonlinear magnetoresistances the magnon creation-annihilation magnetoresistances (\emph{m}$\!^{\dagger}$\!\emph{m}MRs). In particular, the transverse magnetoresistances due to the nonlinear planar Hall effect (\emph{m}$\!^{\dagger}$\!\emph{m}PHE) and nonlinear anomalous Hall effect (\emph{m}$\!^{\dagger}$\!\emph{m}AHE) were previously not accounted for in the harmonic Hall resistance measurements. Due to their similar symmetry to the SOT contributions, the \emph{m}$\!^{\dagger}$\!\emph{m}PHE and \emph{m}$\!^{\dagger}$\!\emph{m}AHE can lead to a strong misestimation of the torques. Consequently, the widespread harmonic Hall resistance measurement technique that has been used to study heavy metals~\cite{garello13, hayashi14, avci14, pai15,lau17, ghosh17, avci17, neumann18,gamou19, du20,shirokura21, ghosh22}, alloys~\cite{nguyen16,  wen17, zhu18, zhu19a, masuda20,hibino20,du21, krishnia21}, oxides~\cite{wang19}, topological insulators~\cite{chi20,binda21,binda23,wang23} and light metals with orbital Hall effect~\cite{lee21, sala22, ding22} should include the \emph{m}$\!^{\dagger}$\!\emph{m}MRs to provide a correct estimation of the torque efficiency. The existence of these nonlinear effects could be the missing piece to understand the discrepancies between the harmonic Hall resistance measurements and other techniques reported in the literature~\cite{lau17, karimeddiny20,zhu20, zhu21, karimeddiny23}.  \\ 

In this work we present a comprehensive approach combining both longitudinal and transverse second harmonic resistance measurements to evaluate the \emph{m}$\!^{\dagger}$\!\emph{m}MRs and SOT contributions in NM/FM bilayers. Accounting for the \emph{m}$\!^{\dagger}$\!\emph{m}MRs, we provide a corrected estimate of the SOTs in different NM/FM material. Specifically, we show that the \emph{m}$\!^{\dagger}$\!\emph{m}PHE, when unaccounted for, can lead to strong misestimation of both the DL and FL SOTs in NM/FM bilayers with in-plane magnetization. To evidence this misestimation, we compare, on the same device, the DL torque estimated by the usual harmonic Hall resistance technique with the one estimated using a calibrated magneto-optical Kerr effect (MOKE) technique. To perform this comparison we used two different conductive FM, Co and Co$_{40}$Fe$_{40}$B$_{20}$ (CoFeB) and two different NM, Pt and W. This comparison reveals a discrepancy of the DL torque, which is overestimated by 15\% in Pt/Co, 30\% in Pt/CoFeB, and up to 100\% in W/CoFeB when using the uncorrected harmonic Hall resistance method. The correction also reveals a misestimation of the magnitude and sign of the FL torque for a rather soft FM like CoFeB. Additionally, we evidence a giant misestimation of both the sign and the order of magnitude of the DL and FL torque in thin Y$_{3}$Fe$_{5}$O$_{12}$/Pt bilayers, where the AHE is much smaller than the PHE and the change of the magnetization due to the magnon creation-annihilation is large. Finally, we comment on how the \emph{m}$\!^{\dagger}$\!\emph{m}PHE can explain discrepancies in previous torque estimations, outline strategies to minimize the torque misestimation in harmonic Hall resistance measurements, and discuss how the \emph{m}$\!^{\dagger}$\!\emph{m}MRs might affect other methods and configurations used to measure SOTs.\\

\section{The harmonic Hall resistance analysis of SOTs} \label{sec:HHR}
\subsection{Standard analysis} \label{sec:HHR_standard}
The change of the magnetization direction due to the SOT in NM/FM bilayers gives rise to a change of the magnetoresistance. As this change is proportional to the current, a second harmonic transverse or longitudinal magnetoresistance can be detected when using an alternating current. It is thus possible to evaluate the SOTs from the transverse magnetoresistance measurement in a Hall bar device ~\cite{pi10, garello13, kim13, avci14, hayashi14}. Thanks to its simplicity, the harmonic Hall resistance analysis has been extensively used to study and evaluate the SOTs in NM/FM bilayers ~\cite{manchon19,nguyen21} both in in-plane magnetized~\cite{avci14, wen17, lau17, neumann18, masuda20,chi20, du20,hibino20, du21,lee21, binda21,binda23} and out-of-plane magnetized samples~\cite{garello13, hayashi14, pai15, nguyen16, ghosh17, avci17, zhu18, zhu19a, wang19, ghosh22, shirokura21, krishnia21, wang23}. More recently it was further used to estimate the SOTs in NM/antiferromagnet bilayers~\cite{cogulu22} as well as the SOTs associated to the orbital Hall effect in light metal/FM bilayers~\cite{sala22, ding22}. The technique further allows for disentangling the current-induced SOT and thermoelectric contributions, as described by Avci \textit{et al.}~\cite{avci14, avci15, ghosh17}. \\

When the magnetization is rotated in-plane by an angle $\varphi$ relative to the current direction, the first and second harmonic transverse magnetoresistance $R_{xy}^{1\omega}$ and $R_{xy}^{2\omega}$ are given by:
\begin{equation}
 R_{xy}^{1\omega}= R_\mathrm{PHE}^{1\omega}\sin\varphi\cos\varphi,
 \label{eq:one}
\end{equation}
\begin{equation}
\begin{split}
 R_{xy}^{2\omega}= (-R_{\mathrm{FL}, xy}^{2\omega}+R_{\mathrm{\nabla T}, xy}^{2\omega}-\frac{1}{2}R_{\mathrm{DL}, xy}^{2\omega})\cos\varphi\\
 + (2R_{\mathrm{FL}, xy}^{2\omega})\cos^{3}\varphi,
 \label{eq:two}
 \end{split}
\end{equation}
where $R_\mathrm{PHE}^{1\omega}$ is the planar Hall resistance, $R_{\mathrm{DL}, xy}^{2\omega}$ is the change of transverse resistance due to the DL torque, $R_{\mathrm{\nabla T}, xy}^{2\omega}$ is the transverse magnetothermal contribution due to the out-of-plane thermal gradient associated with the anomalous Nernst effect (ANE) and spin Seebeck effect (SSE), $R_{\mathrm{FL}, xy}^{2\omega}$ is the change of transverse resistance induced by the Oersted field and the FL torque. For simplicity, we will not separate the FL torque due to the spin accumulation and the Oersted field contributions. The latter can be estimated by Ampere's law or directly measured by NV microscopy, as shown in Sect.~\ref{sec:results_MOKE}. \\

Similarly, the second harmonic longitudinal contribution $ R_{xx}^{2\omega}$ includes contributions from torques and magnetothermal effects, as well as the unidirectional spin Hall magnetoresistance (USMR)~\cite{avci15, langenfeld16, yasuda16, li17, wang18, borisenko18,avci18, sterk19, liu21}. For the in-plane angular dependence, the first and second harmonic longitudinal magnetoresistance contributions $R_{xx}^{1\omega}$ and $R_{xx}^{2\omega}$ are given by:
\begin{equation}
 R_{xx}^{1\omega}= R_\mathrm{AMR}^{1\omega}\cos^{2}\varphi-R_\mathrm{SMR}^{1\omega}\sin^{2}\varphi,
 \label{eq:three}
\end{equation}
\begin{equation}
\begin{split}
    R_{xx}^{2\omega}= (R_\mathrm{USMR}^{2\omega}+R_{\mathrm{\nabla T}, xx}^{2\omega}+2R_{\mathrm{FL}, xx}^{2\omega})\sin\varphi\\
   -2R_{\mathrm{FL}, xx}^{2\omega}\sin^{3}\varphi,
\end{split}
  \label{eq:four} 
\end{equation}
where $R_\mathrm{AMR}^{1\omega}$ is the anisotropic magnetoresistance, $R_\mathrm{SMR}^{1\omega}$ is the spin Hall magnetoresistance, $R_\mathrm{USMR}^{2\omega}$ is the USMR contribution, $R_{\mathrm{\nabla T}, xx}^{2\omega}$ is the longitudinal magnetothermal contribution and $R_{\mathrm{FL}, xx}^{2\omega}$ is the change of the longitudinal resistance due to the FL torque with $R_{\mathrm{FL}, xx}^{2\omega} = -g \cdot R_{\mathrm{FL}, xy}^{2\omega}$ and $R_{\mathrm{\nabla T}, xx}^{2\omega} = -g \cdot R_{\mathrm{\nabla T}, xy}^{2\omega}$, with $g$ the length-to-width ratio of the Hall bar.\\

To disentangle this multitude of effects, one has to rely on their different response to additional external stimuli.  In the standard analysis, Eq.~(\ref{eq:two}) is rewritten as 
\begin{equation}
R_{xy}^{2\omega}= R_{\cos}^{2\omega}\cos\varphi
   + R_{\cos^{3}}^{2\omega}\cos^{3}\varphi,
  \label{eq:two_bis} 
\end{equation}
where $R_{\cos}^{2\omega}=-R_{\mathrm{FL}, xy}^{2\omega}+R_{\mathrm{\nabla T}, xy}^{2\omega}-\frac{1}{2}R_{\mathrm{DL}, xy}^{2\omega}$ and $R_{\cos^{3}}^{2\omega}=2R_{\mathrm{FL}, xy}^{2\omega}$. Assuming that the magnetization aligns with the external field $H$, and by rotating the sample in a constant field, the values of $R_{xy}^{2\omega}$ measured as a function of $\varphi$ are fitted using Eq.~(\ref{eq:two_bis}) to extract the coefficients $R_{\cos}^{2\omega}$ and $R_{\cos^{3}}^{2\omega}$. Using linear combinations of these coefficients, 
\color{black} the torque can be estimated from the variations of $ R_{\mathrm{FL}, xy}^{2\omega}$ and $ R_{\mathrm{DL}, xy}^{2\omega}$ with the applied magnetic field $H$~\cite{avci14}. The FL torque is obtained by:
\begin{equation}
    R_{\mathrm{FL}, xy}^{2\omega} = \frac{1}{2}\frac{R_\mathrm{PHE}^{1\omega}}{\mu_{0}H}B_\mathrm{FL},
      \label{eq:5} 
\end{equation}  
with $\mu_{0}$ the vacuum permeability and $B_\mathrm{FL}$ the magnetic flux density associated to the FL torque. The DL torque is obtained by:
\begin{equation}
    R_{\mathrm{DL}, xy}^{2\omega} + R_{\mathrm{\nabla T}, xy}^{2\omega} = \frac{R_\mathrm{AHE}^{1\omega}}{\mu_{0}H_\mathrm{eff}}B_\mathrm{DL} + R_{\mathrm{\nabla T}, xy}^{2\omega},
      \label{eq:6} 
\end{equation}  
 where the magnetothermal contribution $R_{\mathrm{\nabla T}, xy}^{2\omega}$ does not depend on the effective field $H_\mathrm{eff}$ acting on the magnetization, $R_\mathrm{AHE}^{1\omega}$ is the anomalous Hall resistance and $B_\mathrm{DL}$ the magnetic flux associated to the DL torque. The effective field is given by $H_\mathrm{eff} = H+H_\mathrm{dem}+H_\mathrm{PMA}$ with $H_\mathrm{dem}$ the demagnetizing field and $H_\mathrm{PMA}$ the perpendicular magnetic anisotropy field. In the following, we refer interchangeably to $B_\mathrm{FL}$ as the FL torque and $B_\mathrm{DL}$ as the DL torque.\\

\subsection{Magnonic contribution} \label{sec:HHR_mmMR}

\begin{figure*}[tp] 
\centering 
\includegraphics[width=\textwidth]{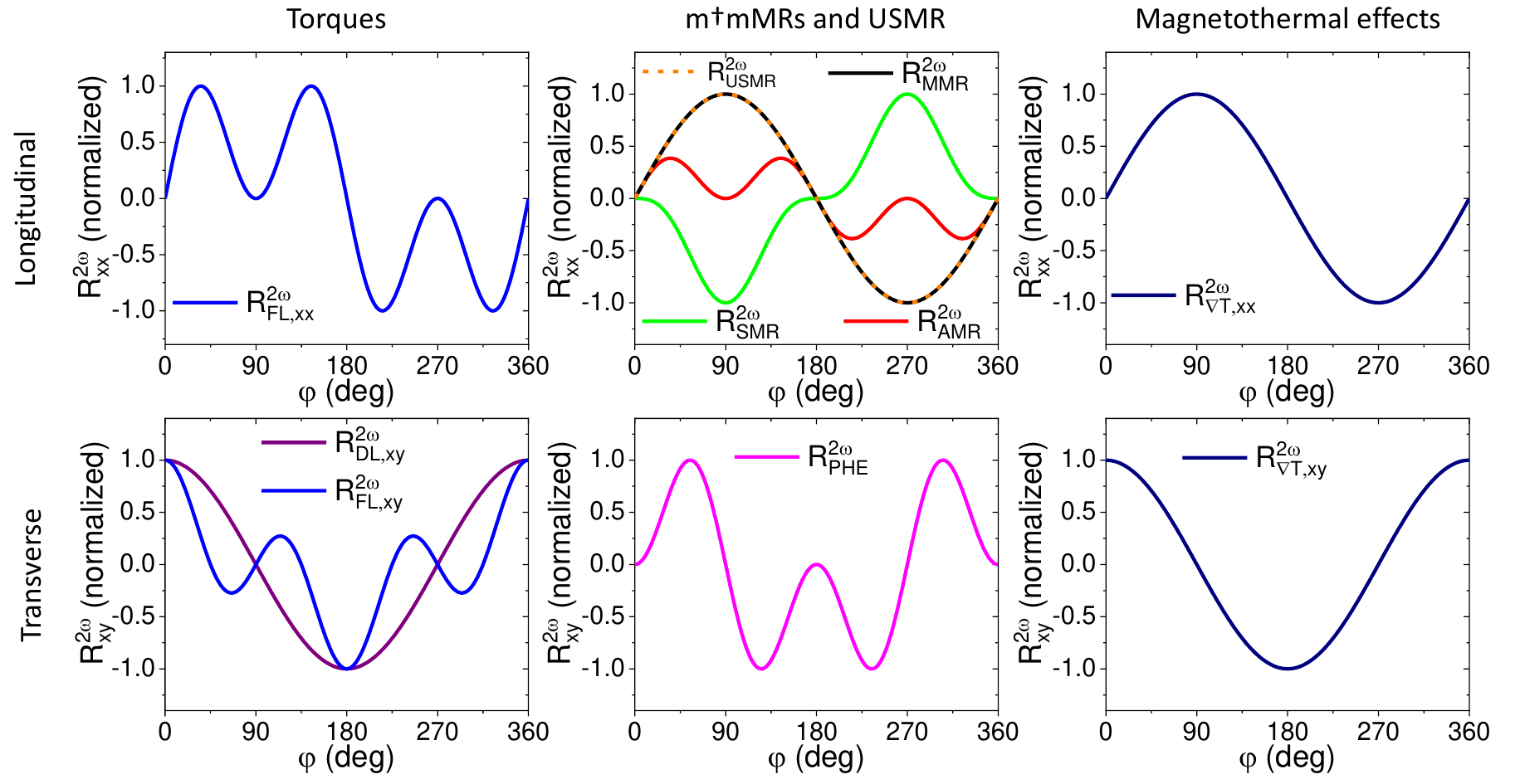}
\caption{\label{fig:figure1} Summary of the angular dependence of the second harmonic signals in the longitudinal and transverse measurement configurations due to the SOTs, \emph{m}$\!^{\dagger}$\!\emph{m}MRs, USMR, and magnetothermal effects.}
\end{figure*}
The analysis of the Hall resistance reported in the previous section is valid when the magnetization $M$ of the FM layer has a constant magnitude, independent of current. However, as discussed in Ref.~\onlinecite{jointpaper}, the injection of a spin current from the NM into the FM layer induces the creation or annihilation of magnons, depending on the polarization of the current. This effect is maximum when the polarization of the spin current is parallel or antiparallel to the magnetization. To first order in the current, the change of magnon population induces a change of magnetization $\Delta M(I)$, resulting in a current- and angle-dependent modulation of $M (I) = M_{s} + \Delta M (I) \sin\varphi$, where $M_{s}$ is the saturation magnetization at the temperature of the measurements. Joule-heating effects $\propto I^2$ can be included in $M_{s}$ to provide higher-order corrections to this expression.\\ 

In order to account for the magnonic contributions Eq.~(\ref{eq:two}) and (\ref{eq:four}) have to be modified. The transverse second harmonic resistance $ R_{xy}^{2\omega}$ and longitudinal second harmonic resistance $ R_{xx}^{2\omega}$ are then given by
\begin{widetext}
\begin{equation}
 R_{xy}^{2\omega}= (R_\mathrm{PHE}^{2\omega}-R_{\mathrm{FL}, xy}^{2\omega}+R_{\mathrm{\nabla T}, xy}^{2\omega}-\frac{1}{2}R_{\mathrm{DL}, xy}^{2\omega})\cos\varphi 
 + (2R_{\mathrm{FL}, xy}^{2\omega}-R_\mathrm{PHE}^{2\omega})\cos^{3}\varphi,
 \label{eq:seven}
\end{equation}
\begin{equation}
    R_{xx}^{2\omega}= (R_\mathrm{USMR}^{2\omega}+R_\mathrm{MMR}^{2\omega}+R_{\mathrm{\nabla T}, xx}^{2\omega}+2R_{\mathrm{FL}, xx}^{2\omega}+R_\mathrm{AMR}^{2\omega})\sin\varphi 
   + (R_\mathrm{SMR}^{2\omega}-R_\mathrm{AMR}^{2\omega}-2R_{\mathrm{FL}, xx}^{2\omega})\sin^{3}\varphi,
  \label{eq:eight} 
\end{equation}
\end{widetext}
where $R_\mathrm{PHE}^{2\omega}$, $R_\mathrm{MMR}^{2\omega}$, $R_\mathrm{AMR}^{2\omega}$ and $R_\mathrm{SMR}^{2\omega}$ are respectively the \emph{m}$\!^{\dagger}$\!\emph{m}PHE, \emph{m}$\!^{\dagger}$\!\emph{m}MMR, \emph{m}$\!^{\dagger}$\!\emph{m}AMR and \emph{m}$\!^{\dagger}$\!\emph{m}SMR coefficients defined in Ref.~\onlinecite{jointpaper}. From Eq.~(\ref{eq:seven}) we notice that $R_\mathrm{PHE}^{2\omega}$ cannot be separated from the $R_{\mathrm{DL}, xy}^{2\omega}$, $R_{\mathrm{FL}, xy}^{2\omega}$ and $R_{\mathrm{\nabla T}, xy}^{2\omega}$ purely based on the angular dependence, since the equation system is underdetermined. Therefore, not accounting for the \emph{m}$\!^{\dagger}$\!\emph{m}PHE using the harmonic Hall resistance method will affect the proper estimation of $R_{\mathrm{DL}, xy}^{2\omega}$, $R_{\mathrm{FL}, xy}^{2\omega}$ and of the associated torques, but will not affect the quality of the fit of $R_{xy}^{2\omega}$. Similar to the transverse resistance, the longitudinal contributions due to the \emph{m}$\!^{\dagger}$\!\emph{m}MRs also cannot be distinguished based on their angular dependence [see Eq.~(\ref{eq:eight})]. We would like to point out that $R_\mathrm{USMR}^{2\omega}$, the USMR contribution, corresponds here to the spin-dependent USMR~\cite{avci15} and the magnonic contribution associated with the spin flip USMR is entirely contained in the expression of $R_\mathrm{MMR}^{2\omega}$. In Fig.~\ref{fig:figure1} we graphically summarized all the possible contributions to the second harmonic signal due to the torques~\cite{avci14}, \emph{m}$\!^{\dagger}$\!\emph{m}MRs~\cite{jointpaper}, USMR~\cite{avci15} and magnetothermal effects~\cite{avci14,miyasato07, uchida08}, in the transverse and longitudinal resistance.

\section{Experiments} \label{sec:exp}

\subsection{Samples} \label{sec:exp_samples}
We performed measurements on Pt(5~nm)/Co$_{40}$Fe$_{40}$B$_{20}$($2.5$~nm), Pt(5~nm)/Co($2.5$~nm), W(5~nm)/Co$_{40}$Fe$_{40}$B$_{20}$($2.5$~nm), W(5~nm)/Co($2.5$~nm), and Y$_{3}$Fe$_{5}$O$_{12}$($6.2$~nm)/Pt(3~nm), where the numbers between parentheses indicate the thickness of each layer. 
The samples with conductive ferromagnets were grown on Si/SiO$_{2}$ substrates by dc magnetron sputtering. The Ar pressure during the growth was $4 \times 10^{-3}$~mbar and the base pressure $1 \times 10^{-7}$~mbar. All samples were capped with a $2.5$-nm-thick Ti layer to avoid the oxidation of the FM in air.  The samples were not annealed. \color{black} Double Hall bars with a length of 100~$\mu$m and a width of 10~$\mu$m (aspect ratio $g \approx 10$) were then patterned using photolithography and Ar etching. To minimize the misestimation of the torques due to inhomogeneous current in the Hall crosses~\cite{garello13,neumann18} we use Hall bars with narrow pick-up lines of 2.5~$\mu$m width. The Y$_{3}$Fe$_{5}$O$_{12}$($6.2$~nm) thin film was deposited using pulsed laser deposition on a (111)-oriented Gd$_{3}$Ga$_{5}$O$_{12}$ substrate and capped with a $3$~nm Pt film by sputtering. The deposition parameters and magnetic properties of the Y$_{3}$Fe$_{5}$O$_{12}$ film are described in Ref.~\cite{mendil19a}. The Pt layer was subsequently patterned into double Hall bars of width 10~$\mu$m and length 50$~\mu$m, with an aspect ratio $g \approx 5$ ~\cite{mendil19b}.

\subsection{Harmonic Hall resistance measurements}  \label{sec:exp_HHR}
The magnetotransport measurement were performed in the double Hall bar devices by injecting an alternating current of frequency $\omega/(2\pi) = \SI{10}{\hertz}$, and using a lock-in detection to measure simultaneously the transverse ($R_{xy}^{1\omega}$, $R_{xy}^{2\omega}$) and longitudinal ($R_{xx}^{1\omega}$, $R_{xx}^{2\omega}$) components of the first and second harmonic resistances. These were obtained by dividing the harmonic components of the measured voltages by the peak amplitude of the current. To vary the angle of the magnetization relative to the current direction, the samples were rotated in an external magnetic field. All the measurements were performed at room temperature. 

\subsection{Magneto optical Kerr effect measurements} \label{sec:exp_MOKE}
In order to provide an independent estimate of the DL torque, we performed complementary measurements using MOKE~\cite{fan14, montazeri15, fan16}. A laser beam with a wavelength of 520~nm was focused onto the Hall bar (spot size of $\approx 0.5$~$\mu$m) and an in-plane magnetic field was applied to orient the magnetization along the bar ($\varphi=0$). An alternating current applied to the Hall bar periodically changes the magnetization direction out of the plane of the sample due to the DL torque and the out-of-plane component of the Oersted field. In this geometry, the magnetization is orthogonal to the polarization of the spin current and the magnonic effects can be safely neglected. The rotation of the light polarization due to this change of the magnetization direction, the Kerr angle $\theta_\mathrm{K}$, was then measured using a balanced photodetector and a lock-in amplifier~\cite{stamm17}. A modulation frequency of $\omega/(2\pi) = 8750$~Hz was chosen for the optical measurement to obtain an optimal signal-to-noise ratio. The incident light was perpendicular to the surface of the sample in the polar configuration. The sample was placed on nanopositioners that allow scanning the laser over its surface. A detailed description of the MOKE setup can be found in Ref.~\onlinecite{stamm17}.\\

\subsection{Scanning nitrogen vacancy magnetometry} \label{sec:exp_NV}
Quantitative MOKE measurements of the DL torque require calibration of the Kerr rotation angle against the effect of a known magnetic field at each value of the current. This is conveniently taken to be the out-of-plane component of the Oersted field, which scales proportionally to the current~\cite{fan14, montazeri15, fan16}. The estimate of the Oersted field typically relies on applying the law of Biot-Savart to an ideal homogeneous conductor with the dimensions of the Hall bar. Here we used scanning nitrogen-vacancy center (NV) magnetometry \cite{Degen2008,Balasubramanian2008} to directly measure the Oersted field emanating from the same Hall bar samples on which the MOKE measurements were performed. A diamond probe containing a single NV center at the apex was scanned at a constant height above the sample surface. The spin resonance of the NV center was excited using a nearby microwave antenna ($2.9$~GHz) and detected by optical readout (532~nm excitation, 630-800~nm detection) readout~\cite{chang17}. The frequency of the NV center's spin resonance is directly proportional to the magnetic field at the probe's apex, which allows for the accurate quantitative mapping of the Oersted field over the Hall bar.

\section{Results}  \label{sec:results}
\subsection{Harmonic Hall resistance measurements of Pt/CoFeB}  \label{sec:results_HHR_CoFeB}
\begin{figure}[tbp]
\includegraphics[width=0.5\textwidth]{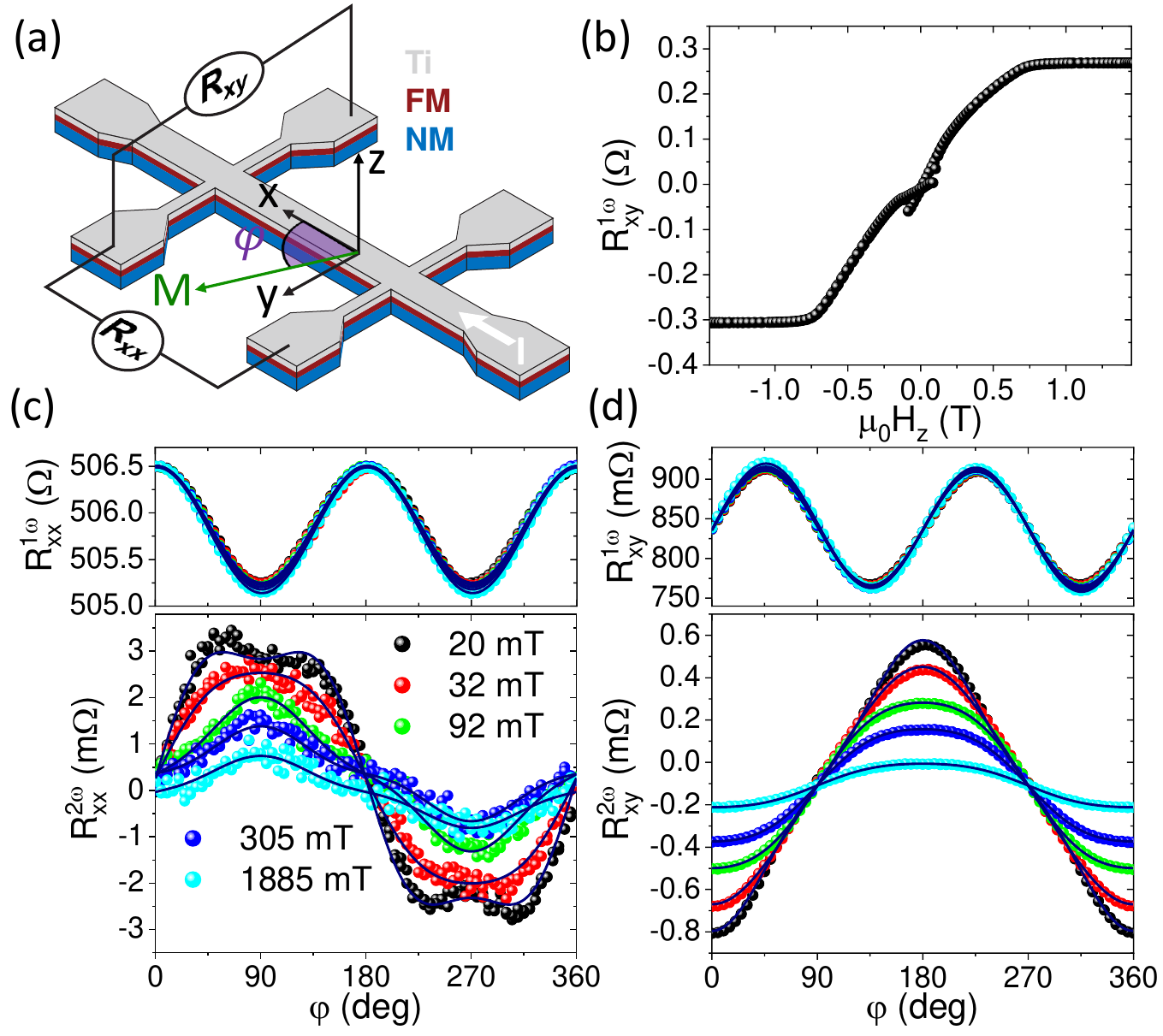}
\caption{\label{fig:figure2}Magnetoresistance measurement in a Pt(5~nm )/CoFeB($2.5$~nm) Hall bar. (a) Experimental setup and coordinate system. The first and second harmonic of the longitudinal $R_{xx}$ and transverse $R_{xy}$ resistances are measured simultaneously. (b) Measurement of the anomalous Hall effect with the magnetic field applied along the $z$ direction. (c) In-plane angular dependence of the first and second harmonic measurement of the transverse magnetoresistance. The lines are the fit to Eq.~(\ref{eq:one}) and (\ref{eq:two}). (d) In-plane angular dependence of the first and second harmonic measurement of the longitudinal magnetoresistance. The lines are the fit to Eq.~(\ref{eq:three}) and (\ref{eq:four}). All measurements were performed at a current of $7.5$~mA.}
\end{figure}

We report an extensive set of measurements performed on a Pt(5~nm)/Co$_{40}$Fe$_{40}$B$_{20}$($2.5$~nm)/Ti($2.5$~nm) double Hall bar. CoFeB is a prototypical FM for spintronics and a standard material for magnetic random access memories~\cite{bhatti17, krizakova22}, while Pt is one of the most studied NM for its high spin-charge interconversion efficiency \cite{garello13, avci14, avci15,manchon19, zhu21}. A schematic of the sample and coordinate system is shown in Fig.~\ref{fig:figure2} (a). First, we measured the anomalous Hall effect with the magnetic field aligned along the $z$ direction as shown in Fig.~\ref{fig:figure2} (b) to evaluate $H_\mathrm{eff}$ and $R_\mathrm{AHE}^{1\omega}$.  In our Pt/CoFeB sample the out-of-plane saturation field is $H_\mathrm{dem}+H_\mathrm{PMA}=710$~mT, which is smaller than the saturation magnetization of 1180 mT due to the large PMA of the Pt/CFB interface. \color{black} We then performed measurements of the in-plane angular dependence of $R_{xy}^{1\omega,2\omega}$ and $R_{xx}^{1\omega,2\omega}$ at different fields and a current of $7.5$~mA as can be seen in Fig.~\ref{fig:figure2} (c) and (d). $R_{xy}^{1\omega}$ exhibits the angular dependence of the PHE described in Eq.~(\ref{eq:one}) while $R_{xx}^{1\omega}$ follows the typical angular dependence of the SMR and AMR given by Eq.~(\ref{eq:three}). Moreover, both $R_{xy}^{2\omega}$ and $R_{xx}^{2\omega}$ are well fitted by Eqs.~(\ref{eq:two}) and (\ref{eq:four}).\\

\begin{figure}[h!]
\includegraphics[width=0.5\textwidth]{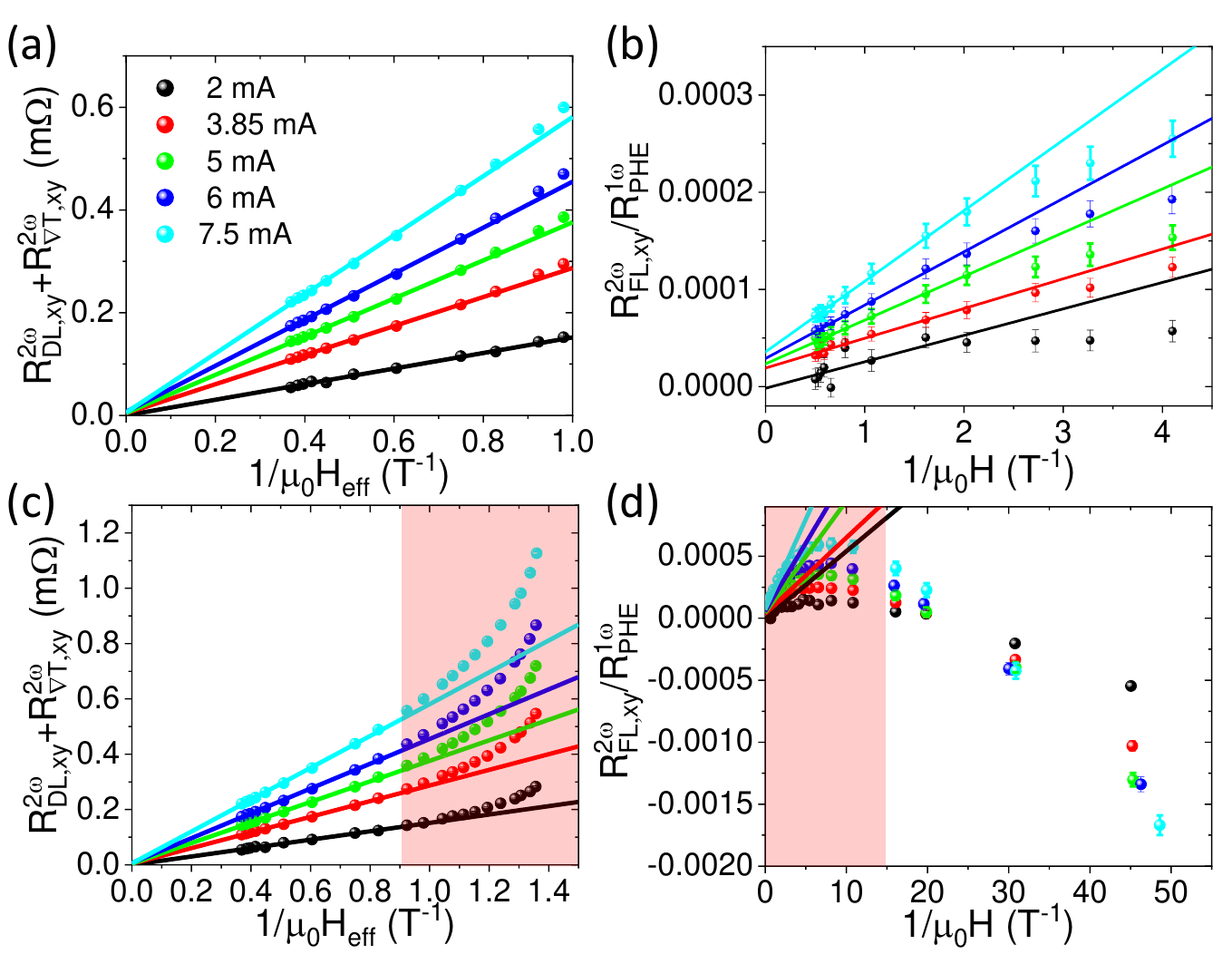}
\caption{\label{fig:figure3} Harmonic Hall resistance analysis in Pt (5~nm)/CoFeB~($2.5$~nm) at different currents. (a) Transverse DL resistance with magnetothermal contribution as a function of $(\mu_0 H_{\rm eff})^{-1}$ and (b) transverse FL resistance divided by the planar Hall resistance as a function of $(\mu_0 H_{\rm eff})^{-1}$. (c) Same as (a) but for fields down to 20~mT (d) Same as (b) but for fields down to 20~mT. The solid lines are linear fits for fields above 250~mT. }
\end{figure}

From the second harmonic Hall resistance, using Eq. \ref{eq:two}, we extracted $R_{\mathrm{DL}, xy}^{2\omega}+R_{\mathrm{\nabla T}, xy}^{2\omega}$ and $R_{\mathrm{FL}, xy}^{2\omega}$ at different magnetic fields and use their field dependence given in Eqs.~(\ref{eq:5}) and (\ref{eq:6}) to obtain an estimate for the SOTs. The results of the harmonic Hall resistance analysis are shown in Fig.~\ref{fig:figure3}. In panels (a) and (b) we show the measurements for magnetic fields above 0.25~T, for both $R_{\mathrm{DL}, xy}^{2\omega}+R_{\mathrm{\nabla T}, xy}^{2\omega}$ and $\frac{R_{\mathrm{FL}, xy}^{2\omega}}{R_\mathrm{PHE}^{1\omega}}$. At high field $R_{\mathrm{DL}, xy}^{2\omega}+R_{\mathrm{\nabla T}, xy}^{2\omega}$ follows the expected dependence $\propto (\mu_0 H_{\rm eff})^{-1}$ and the offset in the limit $\mu_0 H_{\rm eff}\rightarrow \infty$ is close to zero ($R_{\mathrm{\nabla T}} \approx 0$). However, by extending these measurements to fields $\mu_0 H<250$~mT ($\mu_0 H_{\rm eff}< 1$~T), we notice a strong deviation from the linear trend expected from Eq.~(\ref{eq:6}), as seen in Fig.~\ref{fig:figure3} (c). Also the field dependence of $\frac{R_{\mathrm{FL}, xy}^{2\omega}}{R_\mathrm{PHE}^{1\omega}}$ deviates from the expected linear behavior $\propto (\mu_0H)^{-1}$ predicted by Eq.~(\ref{eq:5}) for fields below $0.5$~T and there is an unexpected current-dependent offset for $\mu_0H\rightarrow \infty$. Upon extending the measurements to fields below $0.1$~T, we even observe a sign change of the slope of $\frac{R_{\mathrm{FL}, xy}^{2\omega}}{R_\mathrm{PHE}^{1\omega}}$, as seen in Fig.~\ref{fig:figure3} (d). These deviations and the sign change cannot be explained using the standard harmonic Hall resistance analysis, evidencing the limitations of the model.\\

The deviation from linear behavior of the torque resistances as a function of the inverse of the effective field observed at low field in Fig.~\ref{fig:figure3} (c) and (d) is quite common in measurements of the torque using the harmonic Hall resistance method in NM/FM bilayers~\cite{avci14, lau17, neumann18, mendil19b, gamou19,masuda20, du20}. It was previously associated with the absence of complete saturation of the ferromagnet at low magnetic field~\cite{avci14, lau17} or a uniaxial anisotropy~\cite{neumann18} and only high field values are usually taken into account. However, a non-saturated state or uniaxial anisotropy cannot explain the sign change of the $R_{\mathrm{FL}, xy}^{2\omega}$ and is unlikely to cause nonlinearities up to 250~mT, well above the in-plane saturation field and anisotropy field of our samples. Following the results of Ref.~\onlinecite{jointpaper}, this deviation and sign change are assigned to the \emph{m}$\!^{\dagger}$\!\emph{m}MR associated with the change of magnetization by the creation and annihilation of magnons.\\

\begin{figure}[h!]
\includegraphics[width=0.5\textwidth]{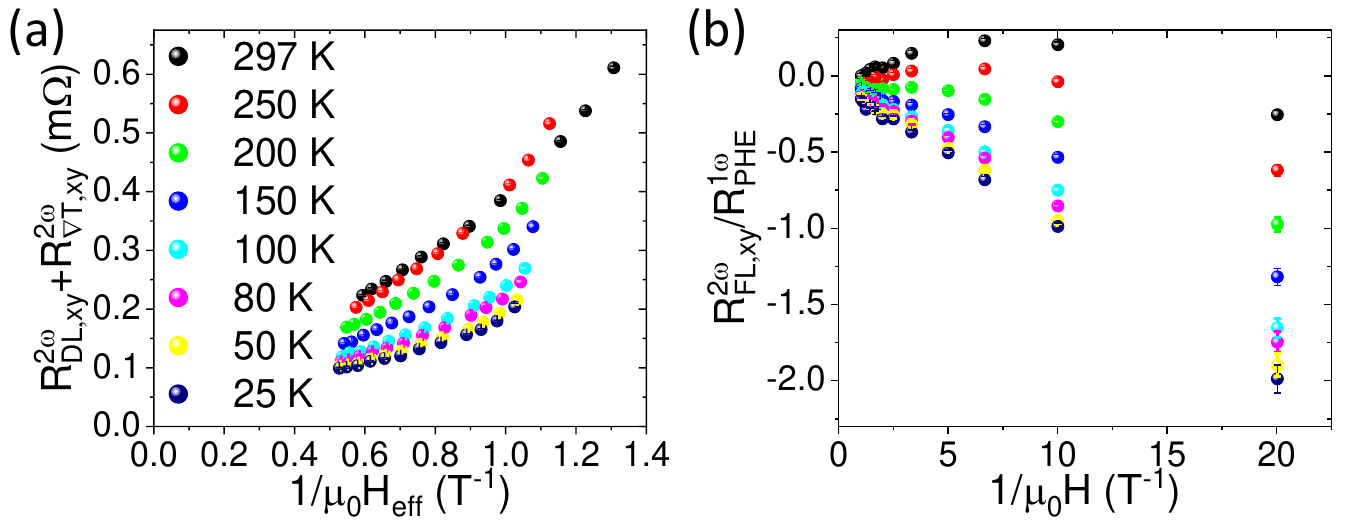}
\caption{\label{fig:figure4} (a) Transverse DL resistance with magnetothermal contribution as a function of $(\mu_0 H_{\rm eff})^{-1}$ and (b) transverse FL resistance divided by
the planar Hall resistance as a function of $(\mu_0 H)^{-1}$ in Pt(5~nm)/CoFeB($2.5$~nm) at a fixed current of 5~mA for different temperatures.}
\end{figure}

To confirm the magnonic origin of the nonlinear behavior of $R_{\mathrm{DL}, xy}^{2\omega}$ and $R_{\mathrm{FL}, xy}^{2\omega}$ observed in Fig.~\ref{fig:figure3}, we performed temperature-dependent harmonic Hall resistance measurements. Due to the lower population of magnons, the magnonic contribution is reduced at low temperature~\cite{goennenwein15, cornelissen16, Li16, avci18}. The in-plane angular dependence measurement of the first and second harmonic resistance was characterized up to a maximum field of 1~T at temperatures between 25~K to 297~K. To minimize the temperature increase of the sample below 3 K and temperature instabilities in the cryostat due to Joule heating, all measurements were performed at a current of 5~mA. The temperature values correspond to the nominal set points of the cryostat, uncorrected for Joule-heating effects. We analyzed the signal using the standard harmonic Hall resistance analysis and performed AHE measurement to evaluate $H_\mathrm{eff}$ and $R_\mathrm{AHE}^{1\omega}$.
The obtained values of $R_{\mathrm{DL}, xy}^{2\omega}+R_{\mathrm{\nabla T}, xy}^{2\omega}$ and $\frac{R_{\mathrm{FL}, xy}^{2\omega}}{R_\mathrm{PHE}^{1\omega}}$ as a function of  $(\mu_0 H_{\rm eff})^{-1}$ and $(\mu_0 H)^{-1}$, respectively, are shown in Fig.~\ref{fig:figure4}. The deviations from the expected linear dependence are reduced when decreasing the temperature for both the DL and FL resistance. This is particularly visible in the field dependence of the FL resistance. Remarkably, the slope is negative in the entire field range for temperature below 200~K, in contrast with the clear change of sign and curvature observed at room temperature. These observations support a contribution of magnonic origin in the second harmonic Hall resistance of the Pt/CoFeB sample.\\

\subsection{MOKE measurements of the DL torque in Pt/CoFeB} \label{sec:results_MOKE}

\begin{figure}[t]
\includegraphics[width=0.45\textwidth]{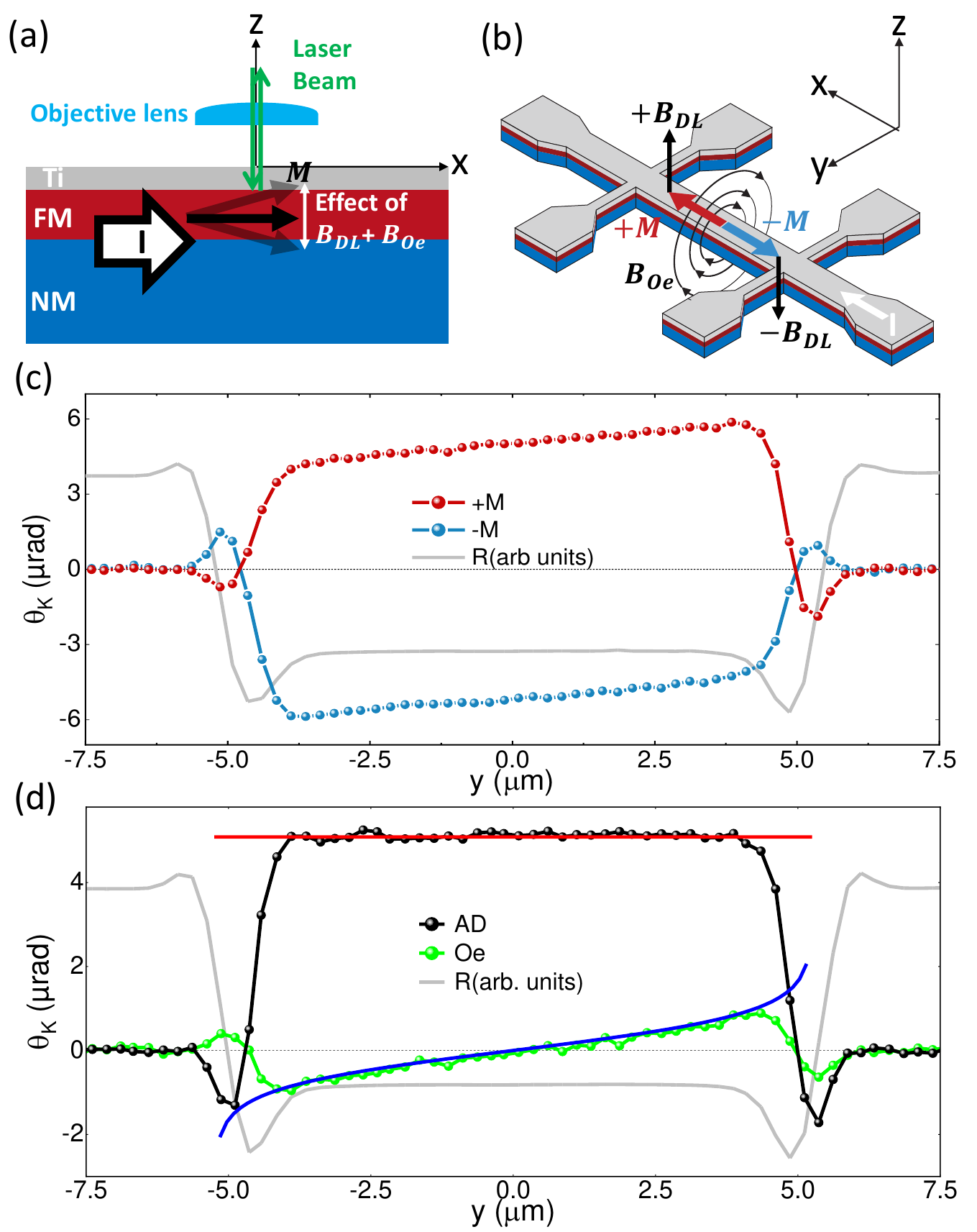}
\caption{\label{fig:figure5}
(a) Schematic of the scanning MOKE microscopy measurements: the effective magnetic field corresponding to the DL torque $B_{\rm DL}$ and the Oersted field $B_{\rm Oe}$ change the direction of the magnetization and thereby the polar MOKE signal. (b) Schematic of the sample and coordinate system. The effect of $B_{\rm DL}$ ($B_{\rm Oe}$) is odd (even) in the magnetization direction. (c) MOKE detection of SOTs in Pt(5~nm)/CoFeB($2.5$~nm): red and blue data points correspond to measurements of the first harmonic Kerr signal at a total current $I_\mathrm{tot} = 3.6$~mA for positive and negative external magnetic field of about 60~mT.  The gray line shows the optical reflectance R of the sample. \color{black} (d) DL and Oersted contribution to the Kerr rotation calculated using Eqs.~(\ref{eq:nine}) and (\ref{eq:ten}). The red line is the linear average of the DL contribution. The blue line is the fit of the out-of-plane Oersted field component calculated using the Biot-Savart law.}
\end{figure}

In order to evaluate the misestimation of the DL torque due to the additional \emph{m}$\!^{\dagger}$\!\emph{m}PHE contribution, we performed a complementary measurement of the DL torque using MOKE. A schematic of the measurements is shown in Fig.~\ref{fig:figure5} (a). The magnetization is aligned along the current direction by a static in-plane magnetic field. Upon applying an ac current, the DL torque causes out-of-plane oscillations of the magnetization that are detected through the first harmonic component of the Kerr rotation angle $\theta_\mathrm{K}$. In addition to the DL torque, the Oersted field produced by the current has perpendicular components near the sample edges, which induce out-of-plane oscillations of the magnetization with opposite sign at opposite edges. 
The respective symmetry of the DL torque and Oersted field allows for separating their contributions~\cite{fan14, montazeri15, fan16}. The Oersted field $B_\mathrm{Oe}$ is only sensitive to the current direction but insensitive to the magnetization direction as it is purely caused by the current. In contrast, the DL field $B_\mathrm{DL}$ is sensitive to both the current direction and magnetization direction as depicted in Fig.~\ref{fig:figure5} (b). Therefore, one can disentangle the Kerr rotation due to the DL torque $\theta_\mathrm{K, DL}$ and Oersted field $\theta_\mathrm{K, Oe}$ using
\begin{eqnarray}
    \theta_\mathrm{K,DL}=\frac{\theta_\mathrm{K}(+M)-\theta_\mathrm{K}(-M)}{2},
    \label{eq:nine}\\ \theta_\mathrm{K, Oe}=\frac{\theta_\mathrm{K}(+M)+\theta_\mathrm{K}(-M)}{2},
    \label{eq:ten}
\end{eqnarray}
where $\pm M$ indicates the direction of the magnetization of the FM layer along $\pm x$. Measurements of the Kerr rotation angle when scanning the laser beam along the $y$ direction obtained in the Pt/CoFeB sample are shown in Fig.~\ref{fig:figure5} (c) for both positive and negative magnetization directions. The Oersted field-induced rotation is asymmetric with respect to the device center, where it vanishes. The DL effective field causes a rotation that is constant across the device~\cite{fan14}. The magnetization direction was changed by applying a magnetic field of about 60~mT along $\pm x$. In Fig.~\ref{fig:figure5} (d) we show the contribution associated with the DL torque and Oersted field obtained using Eqs.~(\ref{eq:nine}) and (\ref{eq:ten}).

The MOKE detection technique offers specific advantages relative to other SOT-measuring techniques. As the magnetization is parallel to the current direction, the signal is inherently free from current-induced magnon creation-annihilation processes. The DL torque in in-plane magnetized samples can be measured by applying a field in the same direction as the current and without the need to rotate the sample or have a field component along $y$, contrary to the second harmonic Hall resistance~\cite{avci14} and spin torque ferromagnetic resonance (ST-FMR) techniques~\cite{liu11, karimeddiny20}. Moreover, the measurement does not require microwave-matched devices as in the case of ST-FMR and can be performed on the very same Hall bar structure as the second harmonic resistance measurements, thus avoiding sample-to-sample variability. The last important point is that the MOKE detection method is calibrated by the Oersted field: estimating the component of the Oersted field along the $z$ direction allows one to determine the proportionality constant between the measured Kerr rotation and the DL effective field. As the film thickness is much smaller than the width of the Hall bar, the effective Oersted field along the $z$ direction, detected by polar MOKE, is given by 
\begin{equation}
    B_\mathrm{Oe}^z(y)=-\frac{\mu_{0}I_\mathrm{tot}}{2 \pi w}  \ln{\frac{w-(y+\frac{w}{2})}{y+\frac{w}{2}}},
    \label{eq:Oe}
\end{equation} 
where $I_\mathrm{tot}$ is the total current in the bilayer, $w$ the width of the Hall bar measured using the reflection of the beam on the device and $y$ the position of the beam with $y = 0$ in the middle of the Hall bar. \\

\begin{figure}[h]
\includegraphics[width=0.4\textwidth]{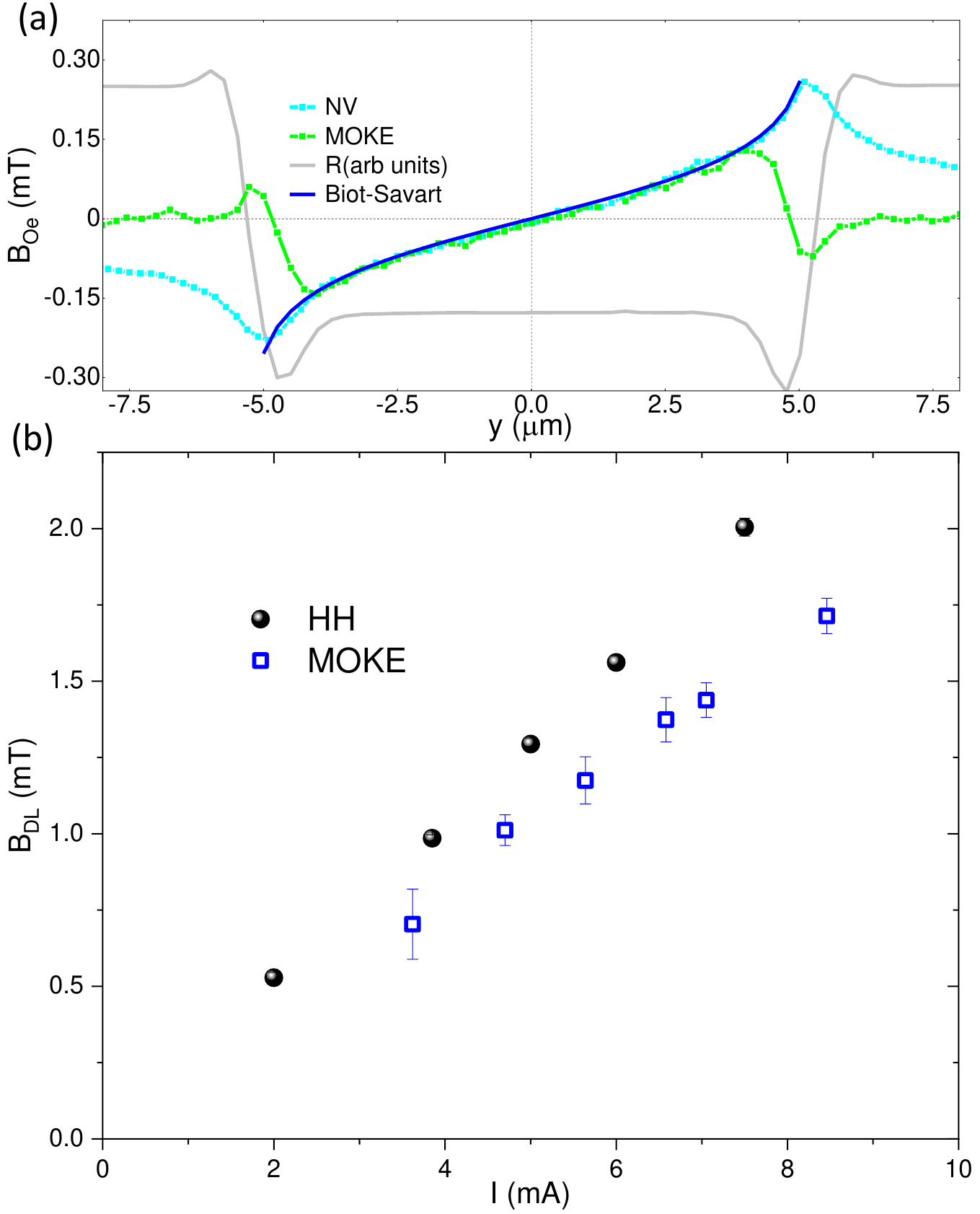}
\caption{\label{fig:figure6} (a) Comparison of the Oersted field measured by MOKE, NV magnetometry and estimated using the Biot-Savart law in Pt(5~nm)/CoFeB($2.5$~nm) at a total current $I_\mathrm{tot} = 3.6$~mA.  The gray line shows the optical reflectance R of the sample.\color{black} (b) Comparison between $B_\mathrm{DL}$ obtained at different currents from MOKE and harmonic Hall resistance (HH) measurements performed on the same device.}
\end{figure}

The proportionality between the Kerr rotation and Oersted field is usually obtained by fitting the Kerr effect contribution to Eq.~(\ref{eq:Oe}) considering that $\theta_\mathrm{K, Oe} \propto B_\mathrm{Oe}^z(y)$. In order to verify the accuracy of the analysis method, we compared the profiles of the Oersted field obtained from the MOKE measurements and Biot-Savart law with independent measurements performed by scanning NV magnetometry on the same sample at the same current. NV magnetometry allows to directly measure the Oersted field due to the current flowing in our Pt/CoFeB bilayer giving a reliable estimate of the exact magnetic field in the vicinity of the device~\cite{chang17}. As can be seen in Fig.~\ref{fig:figure6} (a), the Oersted field profiles obtained from the two methods overlap in the central area of the device, thus supporting the accuracy of our evaluation of the Oersted field and the proportionality between magnetic field and Kerr rotation. Note that MOKE measures the polarization-dependent intensity of the light reflected from the sample's surface and cannot measure the Oersted field outside of the device, whereas NV magnetometry directly measures the Oersted field. Due to stray reflections at the sample edges, and the larger size of the laser beam compared to the tip of the NV microscope, the measurement of the Oersted field using MOKE is not accurate close to the edges.\\

The obtained scaling between the Kerr rotation and field allows for estimating $B_\mathrm{DL}$ for different currents. The value of $B_\mathrm{DL}$ obtained from the harmonic Hall resistance measurement and MOKE on the same device are shown in Fig.~\ref{fig:figure6} (b). The torque obtained using the harmonic Hall resistance method is larger by 30\% than the one obtained optically. This overestimation is associated with the \emph{m}$\!^{\dagger}$\!\emph{m}MRs contribution. We note that the overestimation is opposite to the expected reduction of the torques due to the dispersion of the current in the Hall arms~\cite{garello13,neumann18}, which in our case is expected to be of the order of 5\% due to our narrow pickup lines. Previous works have evidenced discrepancies between the harmonic Hall resistance analysis and other torque measurement methods when using the same samples ~\cite{karimeddiny20,zhu20, zhu21, karimeddiny23}, but an explanation to these discrepancies was lacking. 

\subsection{Harmonic Hall resistance measurements accounting for the \lowercase{\emph{m}$\!^{\dagger}$\!\emph{m}}MR\lowercase{s}} \label{sec:torques_1}

In the following, we propose a method to evaluate the \emph{m}$\!^{\dagger}$\!\emph{m}PHE contribution, correct the estimated values of the DL and FL Hall resistances and compare the newly obtained values of the DL torque to the results of the MOKE detection as a benchmark. In the general case, the contribution of the \emph{m}$\!^{\dagger}$\!\emph{m}PHE cannot be simply obtained using Eq.~(\ref{eq:seven}). Whereas the magnetothermal contribution $R_{\mathrm{\nabla T}, xy}^{2\omega}$ is field-independent and can be obtained at high field when other field-dependent contributions become negligible~\cite{avci14}, the FL, DL, and \emph{m}$\!^{\dagger}$\!\emph{m}PHE contributions cannot be separated by symmetry. In contrast to the FL and DL contributions, the exact field dependence of the \emph{m}$\!^{\dagger}$\!\emph{m}PHE is unknown and depends on the specific magnetic properties of the FM layer, particularly its magnon stiffness~\cite{avci18}. To overcome these limitations we propose to determine the field dependence of the \emph{m}$\!^{\dagger}$\!\emph{m}PHE using the longitudinal \emph{m}$\!^{\dagger}$\!\emph{m}MRs, as the longitudinal and transverse \emph{m}$\!^{\dagger}$\!\emph{m}MRs share the same field dependence \cite{jointpaper}. Our method relies on subtracting the rescaled magnonic contribution from $R_{\mathrm{FL}, xy}^{2\omega}$ and $R_{\mathrm{DL}, xy}^{2\omega}$ until the latter two exhibit the expected field dependent behavior in the absence of magnonic contributions given by Eqs.~(\ref{eq:5}) and (\ref{eq:6}).\\

To obtain the field dependence of the longitudinal \emph{m}$\!^{\dagger}$\!\emph{m}MRs we rewrite Eq.~(\ref{eq:eight}) as
\begin{equation}
R_{xx}^{2\omega}= R_{\sin}^{2\omega}\sin\varphi
   + R_{\sin^{3}}^{2\omega}\sin^{3}\varphi,
  \label{eq:eleven} 
\end{equation}
where $R_{\sin^{3}}^{2\omega} = R_\mathrm{SMR}^{2\omega}-R_\mathrm{AMR}^{2\omega}-2R_{\mathrm{FL}, xx}^{2\omega}$ and $R_{\sin}^{2\omega} = R_\mathrm{USMR}^{2\omega}+R_\mathrm{MMR}^{2\omega}+R_{\mathrm{\nabla T}, xx}^{2\omega}+2R_{\mathrm{FL}, xx}^{2\omega}+R_\mathrm{AMR}^{2\omega}$. We note that the $R_{\mathrm{FL}, xx}^{2\omega}$ term can be eliminated by adding $R_{\sin}$ and $R_{\sin^{3}}$:
\begin{equation}
R_{\sin}^{2\omega} + R_{\sin^{3}}^{2\omega} = R_\mathrm{USMR}^{2\omega}+ R_{\mathrm{\nabla T}, xx}^{2\omega} + R_\mathrm{MMR}^{2\omega} + R_\mathrm{SMR}^{2\omega}.
  \label{eq:twelve} 
\end{equation}
The second harmonic resistances $R_\mathrm{USMR}^{2\omega}$ due to the spin-dependent USMR and $R_{\mathrm{\nabla T}, xx}^{2\omega}$ due to the magnetothermal effects depend only on the magnetization direction and are field-independent~\cite{avci14, avci15} while $R_\mathrm{MMR}^{2\omega}$ and $R_\mathrm{SMR}^{2\omega}$ are magnonic contributions and vary with the external magnetic field. Calculating $R_{\mathrm{magnon}} = R_{\sin}^{2\omega} + R_{\sin^{3}}^{2\omega}$ gives the magnetic field dependence of the magnonic contribution on top of a field-independent offset. Since the \emph{m}$\!^{\dagger}$\!\emph{m}PHE contribution shares the magnonic origin, its magnetic field dependence is expected to be the same as $R_{\mathrm{magnon}}$~\cite{jointpaper}. Equipped with this knowledge, we disentangle the magnonic contributions from the torques in the transverse signal.

\begin{figure}[h]
\includegraphics[width=0.5\textwidth]{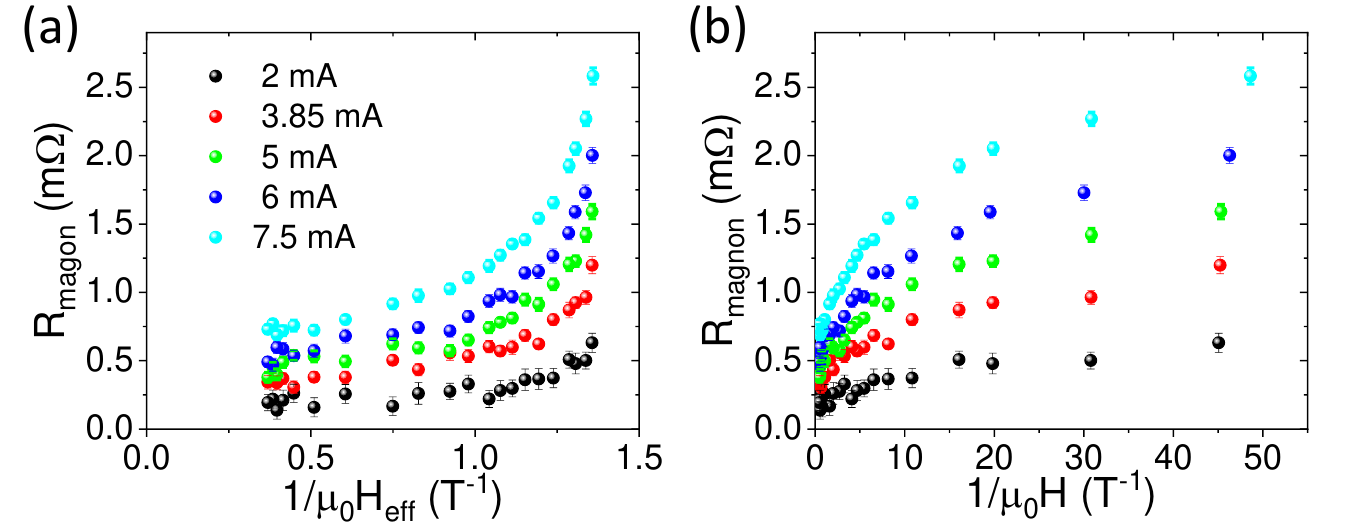}
\caption{\label{fig:figure7} $R_{\mathrm{magnon}}$ as a function of (a) the inverse of the effective magnetic field and (b) the inverse of the magnetic field.}
\end{figure}

To this end we extract $R_{\mathrm{magnon}}$ from the longitudinal resistance measurements in our Pt/CoFeB sample at different currents and fields. $R_{\mathrm{magnon}}$ varies with the external magnetic field and shows the expected finite offset value at high magnetic field due to the USMR and magnetothermal effects. From the results obtained in \ref{sec:results_HHR_CoFeB} the magnonic contribution in Pt/CoFeB is expected to exhibit a field dependence that is similar to the deviation to the expected field dependence of the torque contributions given by Eqs.~(\ref{eq:5}) and (\ref{eq:6}). Indeed, as can be seen in Fig.~\ref{fig:figure7} (a) and Fig.~\ref{fig:figure7} (b) $R_{\mathrm{magnon}}$ as a function of $\frac{1}{\mu_{0}H_\mathrm{eff}}$ and $\frac{1}{\mu_{0}H}$ vary similarly to the deviation from linearity of $R_{\mathrm{DL}, xy}^{2\omega} + R_{\mathrm{\nabla T}, xy}^{2\omega}$ as a function of $\frac{1}{\mu_{0}H_\mathrm{eff}}$ and $\frac{R_{\mathrm{FL}, xy}^{2\omega}}{R_\mathrm{PHE}^{1\omega}}$ as a function of $\frac{1}{\mu_{0}H}$ observed in Fig.~\ref{fig:figure3} (c) and (d). Therefore, $R_{\mathrm{magnon}}$ contains all the nonlinearities that affect the DL and FL signal. \\

As motivated above, the longitudinal magnonic contribution thus provides the field dependence of the \emph{m}$\!^{\dagger}$\!\emph{m}PHE. We then empirically rescale $R_{\mathrm{magnon}}$ to obtain the \emph{m}$\!^{\dagger}$\!\emph{m}PHE contribution using
\begin{equation}
    R_\mathrm{PHE}^{2\omega} = C_\mathrm{mag}(R_{\mathrm{magnon}}-R_{\mathrm{off}}),
    \label{eq:thirteen}
\end{equation}
where $C_\mathrm{mag}$ is a scaling factor that depends on the aspect ratio and magnetoresistive properties of the device and $R_{\mathrm{off}}$ the offset constant due to the spin-dependent USMR and magnetothermal contributions. In the following we take $R_{\mathrm{off}}$ to be equal to $R_{\mathrm{magnon}}$ at the maximum applied field, for which the field-dependent magnonic contribution is the lowest. \\

\begin{figure}[h]
\includegraphics[width=0.5\textwidth]{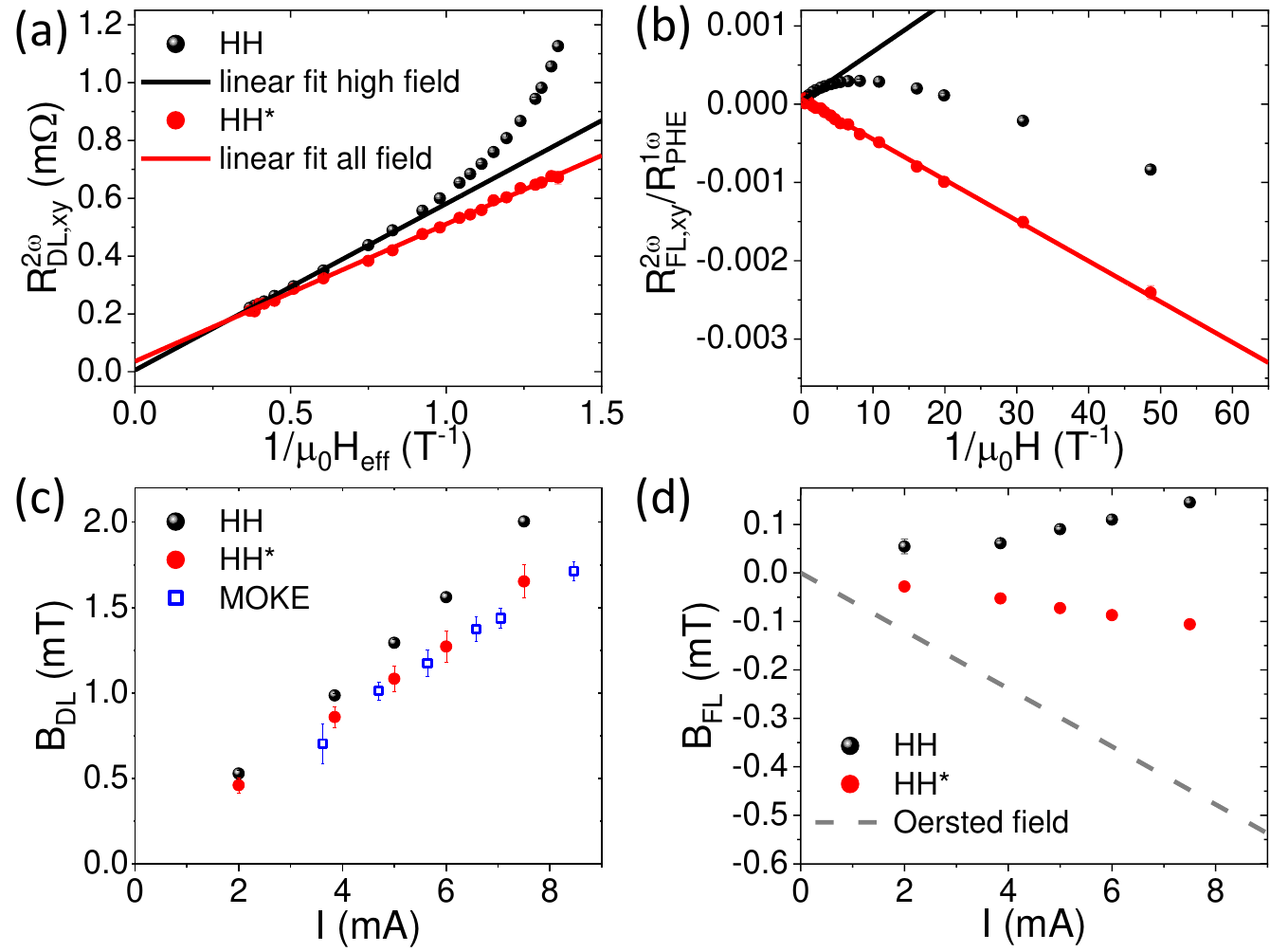}
\caption{\label{fig:figure8} Correction to the harmonic Hall resistance measurements in Pt(5~nm)/CoFeB($2.5$~nm). (a) Transverse DL resistance  with magnetothermal contribution at a current of $7.5$~mA using the standard harmonic Hall resistance analysis and after correction using Eq.~(\ref{eq:fourteen}). (b) Transverse FL resistance divided by the planar Hall resistance at a current of $7.5$~mA using the standard harmonic Hall resistance analysis and after correction using Eq.~(\ref{eq:fifteen}). (c) Comparison between the DL torque obtained at different currents from MOKE and the harmonic Hall resistance analysis with and without magnon correction. (d) Comparison between the FL torque obtained at different currents from the harmonic Hall resistance analysis with and without magnon correction. The estimated Oersted field is plotted as a dashed line.}
\end{figure}

We can now account for the \emph{m}$\!^{\dagger}$\!\emph{m}PHE contribution in the torque estimation using Eq.~(\ref{eq:thirteen}) and obtain the corrected DL and FL contributions
\begin{eqnarray}
    R_{\mathrm{DLcorr}, xy}^{2\omega} = R_{\mathrm{DL}, xy}^{2\omega}-C_\mathrm{mag}(R_{\mathrm{magnon}}-R_{\mathrm{off}})
    \label{eq:fourteen},\\
    R_{\mathrm{FLcorr}, xy}^{2\omega} = R_{\mathrm{FL}, xy}^{2\omega} - \frac{C_\mathrm{mag}}{2}(R_{\mathrm{magnon}}-R_{\mathrm{off}})
    \label{eq:fifteen},
\end{eqnarray}
where the value of $C_\mathrm{mag}$ is obtained through minimizing the residuals of the linear fit of $R_{\mathrm{DLcorr}, xy}^{2\omega}$ and $\frac{R_{\mathrm{FLcorr}, xy}^{2\omega}}{R_\mathrm{PHE}^{1\omega}}$ as a function of the inverse of $(\mu_0 H_{\rm eff})^{-1}$ and $(\mu_0 H)^{-1}$, respectively. In the following, we will present the results obtained using the standard harmonic Hall resistance analysis (HH) and  compare them with the corrected ones (HH*) and the MOKE detection. The corrected values of the transverse DL and FL resistances at a current of $7.5$~mA with $C_\mathrm{mag}$ = 0.24 $\pm$ 0.02 are shown in Fig.~\ref{fig:figure8} (a) and (b). A similar value of $C_\mathrm{mag}$ was obtained at all currents in our Pt/CoFeB device. Both $R_{\mathrm{DLcorr}, xy}^{2\omega}$ and $\frac{R_{\mathrm{FLcorr}, xy}^{2\omega}}{R_\mathrm{PHE}^{1\omega}}$ are linear for the same value of $C_\mathrm{mag}$, which supports the validity of the magnon correction. \\

With the correction, we find a reduced DL torque due to the decrease of the slope even in the high magnetic field range above 250~mT. As can be seen in Fig.~\ref{fig:figure8} (c), the values obtained using the corrected harmonic Hall resistance measurement are in much better agreement with the MOKE. For the total FL torque, including the Oersted field contribution, the result is even more striking. Without the correction, the sign of the total FL torque is positive, opposite to the Oersted field, which is negative for this stacking order. After correction, the sign of the total FL torque is reversed and the same as the torque due to the Oersted field, as seen in Fig.~\ref{fig:figure8} (d). 
The total Oersted field contribution is estimated using $B_\mathrm{Oe}=-\frac{\mu_{0}I_\mathrm{NM}}{2 w}$, where $I_\mathrm{NM}$ is the current flowing in the NM layer, evaluated using a parallel resistor model. Note that $B_\mathrm{Oe}$ here refers to the in-plane component of the Oersted field, not the spatially inhomogeneous out-of-plane component measured by MOKE in Sect.~\ref{sec:results_MOKE}. Due to the high resistivity of the 2.5~nm CoFeB layer ($\rho_{\mathrm{CoFeB}}$ = 260 $\mu \Omega$ cm) compared to the resistivity of Pt ($\rho_{\mathrm{Pt}}$ = 27.5 $\mu \Omega$ cm), 95\% of the applied current flows in the Pt layer. In the Pt/CoFeB sample studied here, the FL torque after subtraction of the Oersted contribution would be overestimated by a factor of 2 without accounting for the \emph{m}$\!^{\dagger}$\!\emph{m}PHE. This result is of particular importance as the FL torque is usually considered as a signature of the interfacial spin-orbit coupling and as a hallmark of the Rashba-Edelstein effect ~\cite{gambardella11,haney13,manchon19, bonell20}. Such a misestimation of the FL torque could thus lead to erroneous interpretations on the role of the interface in the SOTs. 

The correction described above provides an accurate estimate of both the DL and FL torques measured by the harmonic Hall resistance method in the presence of magnonic excitations. The latter are expected to be relevant in relatively "soft" magnetic layers, such as CoFeB, and at high current densities. In these cases, as we will further elaborate in Sects.~\ref{sec:torques_2} and \ref{sec:YIG}, the DL torque can be largely overestimated and the FL torque may  appear with the wrong sign. We note also that the correction of the Hall resistance may increase the error bars of the torques relative to the uncorrected measurements. The larger error bars of the DL torque originate from the larger noise of the second harmonic longitudinal resistance compared to the transverse resistance, as well as the uncertainty of $C_\mathrm{mag}$. For the FL torque, the main source of error in the uncorrected data is the deviation from linearity of $\frac{R_{\mathrm{FL}, xy}^{2\omega}}{R_\mathrm{PHE}^{1\omega}}$ as a function of $(\mu_0H)^{-1}$. The correction improves the fit and therefore does not strongly increase the error. \\

We conclude this section by summarizing the protocol for conducting SOT measurements on samples with in-plane magnetization taking into account the magnon contributions:

\begin{itemize}
    \item[(i)] Measure the angular dependence of the longitudinal and transverse first and second harmonic resistances $R_{xx}^{1\omega,2\omega}$ and $R_{xy}^{1\omega,2\omega}$ at different external magnetic fields over a broad field range. 

  \item[(ii)] For each field, fit $R_{xy}^{2\omega}= R_{\cos}^{2\omega}\cos\varphi + R_{\cos^{3}}^{2\omega}\cos^{3}\varphi$ (Eq.~\ref{eq:two_bis}) and $R_{xx}^{2\omega}= R_{\sin}^{2\omega}\sin\varphi + R_{\sin^{3}}^{2\omega}\sin^{3}\varphi$ (Eq.~\ref{eq:eleven}) to extract the coefficients of the different trigonometric functions.
  
  \item[(iii)]  Using the standard analysis (Eqs.~\ref{eq:5} and \ref{eq:6}), for each field, calculate $R_{\mathrm{\nabla T}, xy}^{2\omega} -\frac{1}{2}R_{\mathrm{DL}, xy}^{2\omega}= R_{\cos}^{2\omega} + \frac{1}{2}R_{\cos^{3}}^{2\omega}$ and $R_{\mathrm{FL}, xy}^{2\omega}= \frac{1}{2}R_{\cos^{3}}^{2\omega}$.

    \item[(iv)] Extract $R_{\mathrm{magnon}}=R_{\sin}^{2\omega} + R_{\sin^{3}}^{2\omega}$ from the fits of $R_{xx}^{2\omega}$ at each field. 
     
   If the field dependence of $R_{\mathrm{DL}, xy}^{2\omega}$, $R_{\mathrm{\nabla T}, xy}^{2\omega}$ and $R_{\mathrm{FL}, xy}^{2\omega}$ is in good agreement with the one given in Eqs. \ref{eq:5} and \ref{eq:6}, and if $R_{\mathrm{magnon}}$ shows no field dependence, no correction is needed. The SOTs can be extracted using the standard harmonic Hall method described in Ref.~\cite{avci14}. 

    If the field dependence of $R_{\mathrm{DL}, xy}^{2\omega}$, $R_{\mathrm{\nabla T}, xy}^{2\omega}$ and $R_{\mathrm{FL}, xy}^{2\omega}$ deviates from the one given in Eqs. \ref{eq:5} and \ref{eq:6}, or if $R_{\mathrm{magnon}}$ is field dependent, the magnonic contribution must be subtracted to obtain a proper estimate of the SOTs.

    \item[(v)] Subtract the magnon contribution using Eqs. \ref{eq:fifteen} and \ref{eq:sixteen}. The corrected values of the DL and FL contributions are obtained by iterative approximations until their field dependence does not deviate from the expected one given in Eqs. \ref{eq:5} and \ref{eq:6}. 

    \item[(vi)] Use the corrected $ R_{\mathrm{DLcorr}, xy}^{2\omega}$ and $ R_{\mathrm{FLcorr}, xy}^{2\omega}$ to estimate the DL and FL torques using Eqs. \ref{eq:5} and \ref{eq:6}.
\end{itemize}
\color{black}

\subsection{Temperature dependence of the torques in Pt/CoFeB accounting for the \lowercase{\emph{m}$\!^{\dagger}$\!\emph{m}}MR\lowercase{s}} \label{sec:torques_1_Tdep}
\begin{figure}[h!]
\includegraphics[width=0.5\textwidth]{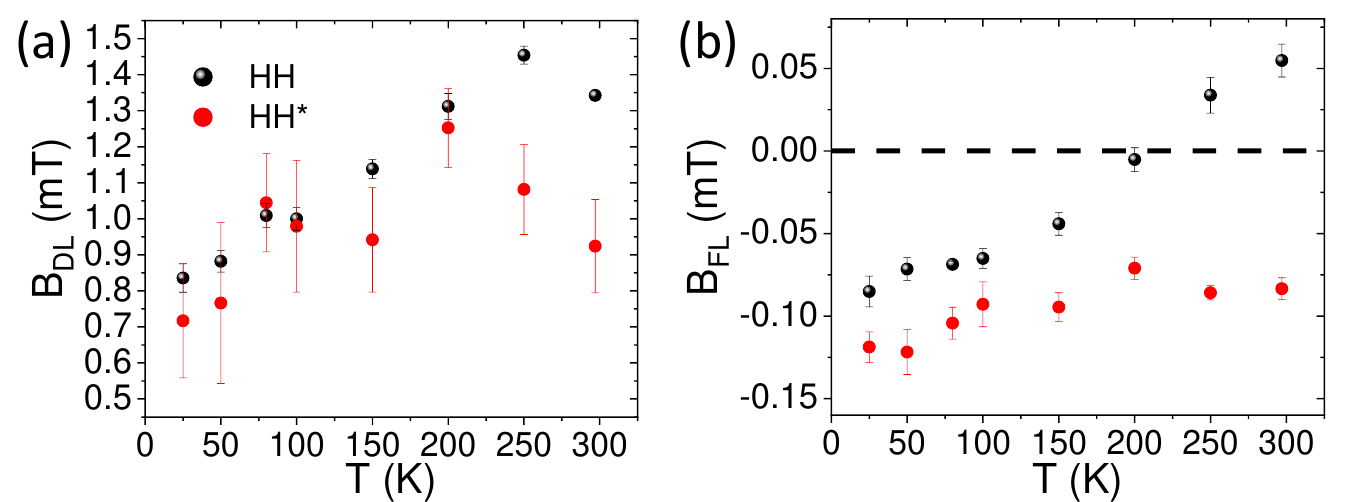}
\caption{\label{fig:figure9} Temperature dependence of the DL and FL torques in Pt(5~nm)/CoFeB($2.5$~nm) at a fixed current of 5~mA (a) DL torque measured using the standard harmonic Hall resistance analysis and with corrections using Eq.~(\ref{eq:fourteen}). (b) FL torque measured using the standard harmonic Hall resistance analysis and with corrections using Eq.~(\ref{eq:fifteen}).}
\end{figure}

Using the harmonic Hall resistance data reported in Fig.~\ref{fig:figure4} we can now estimate the DL and FL torques as a function of temperature with and without the \emph{m}$\!^{\dagger}$\!\emph{m}PHE correction. The correction of the magnon effects is performed by taking the scaling factor $C_\mathrm{mag}$ = 0.24 $\pm$ 0.02 estimated at room temperature to be the same at all temperatures. Figures~\ref{fig:figure9} (a) and (b) show the corrected and uncorrected FL and DL torques measured from 25~K to room temperature. The discrepancy between the corrected and uncorrected FL and DL torques is largest for temperatures above 150~K. Below 150~K, when the magnon contribution is reduced, the measurements using the second harmonic Hall resistance without correction are in much better agreement with the corrected estimation for both the FL and DL torques. \\

In the uncorrected measurements, both the DL torque and the FL torque appear to have a strong temperature dependence. The total FL torque even changes sign at around 200~K. However, when accounting for the \emph{m}$\!^{\dagger}$\!\emph{m}PHE, the temperature dependence of the DL and FL torques is strongly modified. In particular, the corrected FL torque does not change sign as a function of temperature, in striking contrast with the uncorrected measurement. In the corrected data, we observe only a small change of around 20\% of the total FL torque as a function of temperature. The temperature dependence of the DL torque is also reduced compared to the uncorrected measurements. The change of the magnonic part of the harmonic Hall signal can thus give rise to an apparent dependence of the torques on temperature, originating from the freeze-out of magnons. Controlling the possible contribution of the \emph{m}$\!^{\dagger}$\!\emph{m}PHE is thus essential to accurately evaluate the temperature dependence of the torques and better understand their physical origin~\cite{pai15, wen17, peterson18}.

\subsection{Estimation of the torques in Pt/Co, W/CoFeB, and W/Co} \label{sec:torques_2}

To confirm that the contribution of the \emph{m}$\!^{\dagger}$\!\emph{m}PHE applies to all NM/FM bilayers, we performed a comprehensive set of harmonic Hall resistance measurements at room temperature on Pt (5~nm)/ Co ($2.5$~nm), W (5~nm)/Co$_{40}$Fe$_{40}$B$_{20}$ ($2.5$~nm), and W(5~nm)/Co($2.5$~nm), all capped by a $2.5$-nm-thick Ti layer. The W layer used in this study has a resistivity $\rho_{\mathrm{W}}$ = 110 $\mu \Omega$ cm comparable to the one of $\beta$-W known for its high spin Hall angle~\cite{pai12, ghosh22}.\\

\begin{figure}[h]
\includegraphics[width=0.5\textwidth]{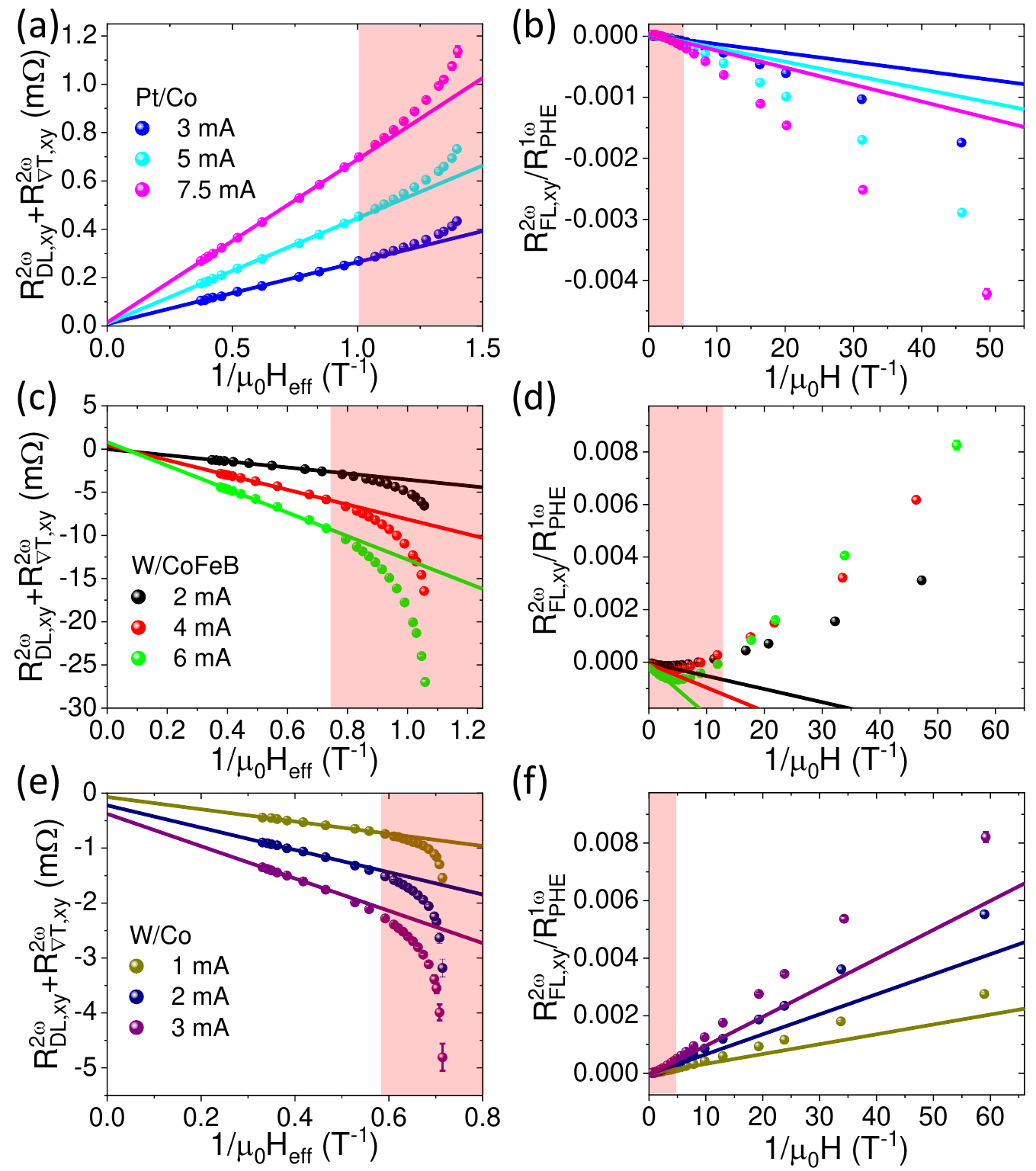}
\caption{\label{fig:figure10} Harmonic Hall resistance analysis of the DL and FL torques in different NM/FM bilayers at different currents. (a) Transverse DL resistance  with magnetothermal contribution and (b) transverse FL resistance divided by the planar Hall effect in Pt (5~nm)/ Co ($2.5$~nm). (c) and (d) are the same for W(5~nm)/CoFeB($2.5$~nm) and (e) and (f) for W(5~nm)/Co($2.5$~nm). The solid lines are linear fits performed for fields larger than 250~mT.}
\end{figure}

Figure~\ref{fig:figure10} reports the plots of $R_{\mathrm{DL}, xy}^{2\omega}+R_{\mathrm{\nabla T}, xy}^{2\omega}$ and $\frac{R_{\mathrm{FL}, xy}^{2\omega}}{R_\mathrm{PHE}^{1\omega}}$ as a function of $(\mu_0 H_{\rm eff})^{-1}$ and $(\mu_0 H^{-1})$, respectively, obtained for these samples. A consistent deviation from linearity is observed for all samples for both the DL torque and FL torque resistances, evidencing the presence of the \emph{m}$\!^{\dagger}$\!\emph{m}PHE contribution in several commonly-used NM/FM bilayers. The amplitude of the deviation varies with both the NM and FM used and will thus affect the torque estimations using the harmonic Hall resistance analysis in different proportions.

\begin{figure}[h]
\includegraphics[width=0.5\textwidth]{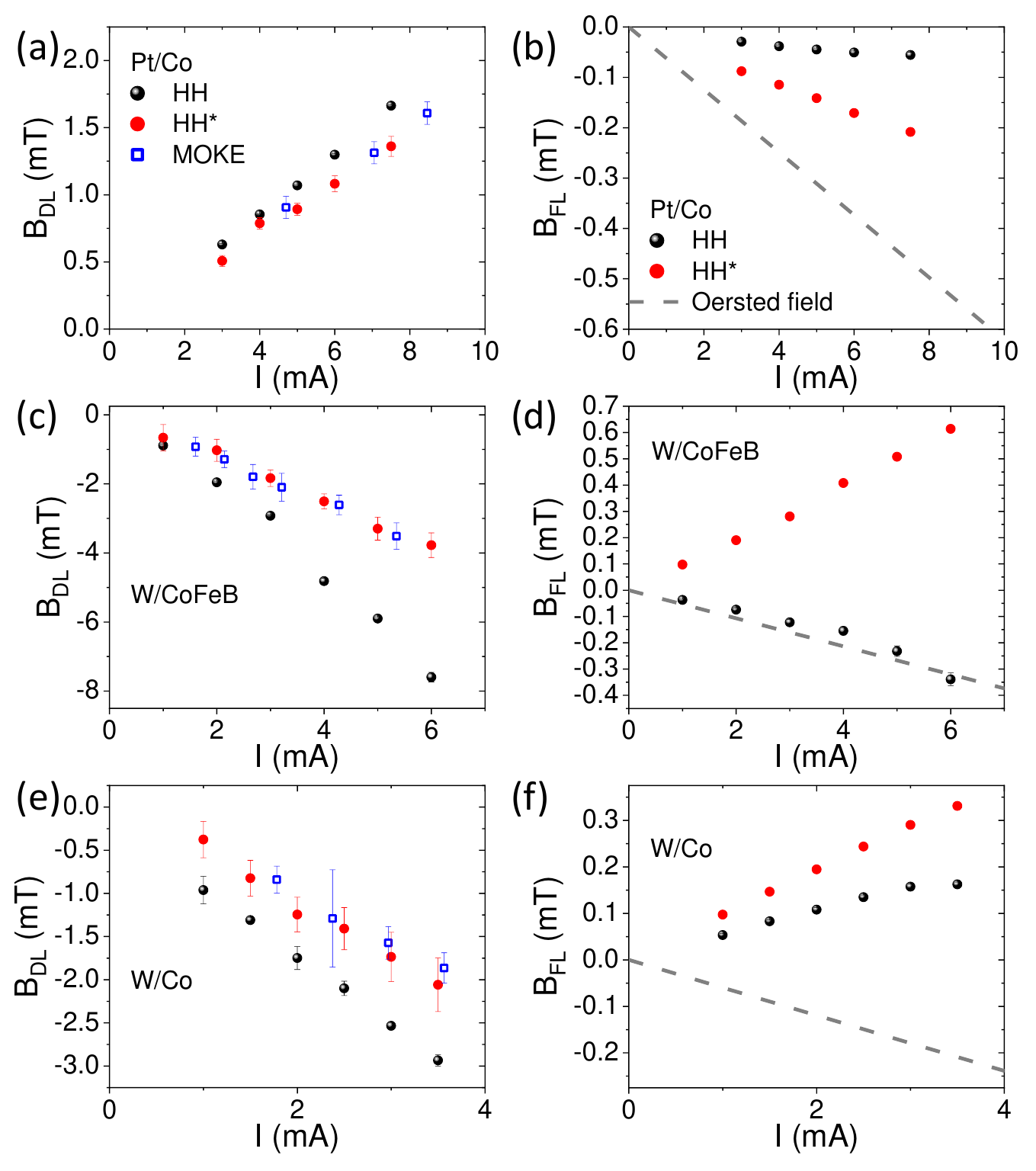}
\caption{\label{fig:figure11} (a) DL torque extracted from MOKE and harmonic Hall resistance measurements with and without correction and (b) FL torque extracted from harmonic Hall resistance measurement with and without correction in Pt (5~nm)/ Co ($2.5$~nm). (c) and (d) are the same for W(5~nm)/CoFeB($2.5$~nm) and (e) and (f) for W(5~nm)/Co($2.5$~nm). The estimated Oersted field is plotted as a dashed line.}
\end{figure}

The DL and FL torques estimated using the standard harmonic Hall resistance analysis described in Sect.~\ref{sec:HHR_standard} and by correcting for the magnonic contributions according to the procedure outlined in Sect.~\ref{sec:torques_1} are shown in Fig.~\ref{fig:figure11}. To provide an independent benchmark, we also measured the DL torque in the Pt/Co, W/CoFeB and W/Co samples using the MOKE technique. Similarly to the case of Pt/CoFeB, we find excellent agreement between the torques measured using the corrected harmonic Hall resistance method and MOKE. When not accounting for the \emph{m}$\!^{\dagger}$\!\emph{m}PHE (see Fig.~\ref{fig:figure11}), the overestimation of the DL torque is around 15\% for Pt/Co and 45\% in W/Co and can be up to 100\% in W/CoFeB. We note that the overestimation qualitatively scales with the amount of deviation from linearity observed in Fig.~\ref{fig:figure10}.

The scaling factors $C_\mathrm{mag}$ for both the DL and FL torque are 0.15  $\pm$ 0.02 in Pt/Co and 0.17 $\pm$ 0.02 in W/Co. In the W/CoFeB case, we find $C_\mathrm{mag}$ = 0.26 $\pm$ 0.01 for the FL torque. Using this value also for the DL torque does not completely cancel the nonlinearity at low field for the transverse DL resistance but still result in a good agreement of the DL torque with the MOKE measurements. In W/CoFeB the magnonic contribution and torque misestimation is larger than in the other samples, and the correction is more sensitive to the estimate of $R_{\mathrm{magnon}}$ measured on the noisier longitudinal channel. \\

We notice two important characteristics regarding our series of samples. First, the deviation from linearity and misestimation of the DL torque is larger when the NM is W instead of Pt. In our W/FM samples the PHE is comparable to the AHE while it is half of the AHE in Pt/FM samples. This is due to the larger spin Hall angle in W and the consequently increased SMR and PHE~\cite{cho15,kim16,lau17}. As the \emph{m}$\!^{\dagger}$\!\emph{m}PHE scales with the PHE and the DL resistance with the AHE, this inevitably leads to a larger correction in W. The larger spin Hall angle of W compared to Pt also leads to a larger change in the total magnon population for the same current density and thus a larger misestimation. Second, the deviation is smaller when the FM is Co instead of CoFeB. The relative weight of AHE and PHE is comparable in both FM and cannot explain the lower contribution of the \emph{m}$\!^{\dagger}$\!\emph{m}PHE in Co. The magnetic damping of Co, however, is usually larger than in CoFe alloys~\cite{schoen16, schoen17}. Moreover, the Curie temperature of Co is larger compared with CoFeB~\cite{yamanouchi11}. Both effects point towards smaller magnon creation-annihilation effects in Co relative to CoFeB.  More generally, the magnon stiffness of the FM is expected to affect the total change of the magnon population as a function of the external magnetic field and current~\cite{rowan-robinson14,avci18}. 

We observe a similar trend for the FL torque. The correction due to the \emph{m}$\!^{\dagger}$\!\emph{m}PHE is again larger when the FM is CoFeB and the NM is W, as can be seen in Fig.~\ref{fig:figure11} (b), (d), (f). When the FM is CoFeB the sign of the total FL torque, including the Oersted field, is also misestimated. In the case of W/CoFeB the FL torque obtained using the standard harmonic Hall resistance analysis is comparable to the Oersted field, but is actually of opposite sign and two times larger when correcting for the magnonic contribution. The sign change in the total FL torque between NM/Co and NM/CoFeB is suppressed when correcting the \emph{m}$\!^{\dagger}$\!\emph{m}MRs contribution and both show similar amplitudes at comparable current densities. 

We conclude that the contribution of the \emph{m}$\!^{\dagger}$\!\emph{m}MRs depends both on the properties of the FM and of the NM. We discuss the ideal characteristics of the FM to avoid a large misestimation of the torques using the standard Hall resistance analysis in Sect.~\ref{sec:mitigation}. \\

 \subsection{Giant torque misestimation in Y$_{3}$F$\mathrm{e}$$_{5}$O$_{12}$/P\lowercase{t}}  \label{sec:YIG}
In YIG, the damping and the magnetization are typically one order of magnitude smaller than in conductive FM \cite{onbasli14,beaulieu18}. The Curie temperature $T_\mathrm{c}$ = 560~K in bulk YIG \cite{cherepanov93} is also smaller compared with the usual metallic FM in spintronics devices leading to a larger total population of magnons at room temperature. For these reasons, YIG is considered as an ideal material platform to study magnonic effects~\cite{goennenwein15, cornelissen16, Li16,demidov17,thiery18}. Therefore, for similar spin accumulation at the interface and comparable thickness of the FM layer, the relative change of the magnon population is expected to be larger in YIG. We thus expect that the spin-orbit torques estimated using the standard harmonic Hall resistance analysis may be largely overestimated.\\

To explore the consequences of the \emph{m}$\!^{\dagger}$\!\emph{m}MRs for torque estimation also in FM/NM bilayers with insulating FM, we performed an in-plane angular dependent measurement of the first and second harmonic longitudinal and transverse magnetoresistance at different external magnetic field in Y$_{3}$Fe$_{5}$O$_{12}$(6.2~nm)/Pt(3~nm) (YIG/Pt). The measurements on this sample are also discussed in detail in our joint paper and the reader is directed there for the corresponding discussion~\cite{jointpaper}. We focus here on the comparison between the conventional harmonic Hall resistance analysis, which assumes $R_\mathrm{PHE}^{2\omega} = 0$ and the estimation of torques when accounting for the \emph{m}$\!^{\dagger}$\!\emph{m}MRs using the method described in Sect.~\ref{sec:torques_1}.\\

\begin{figure}[h]
\includegraphics[width=0.5\textwidth]{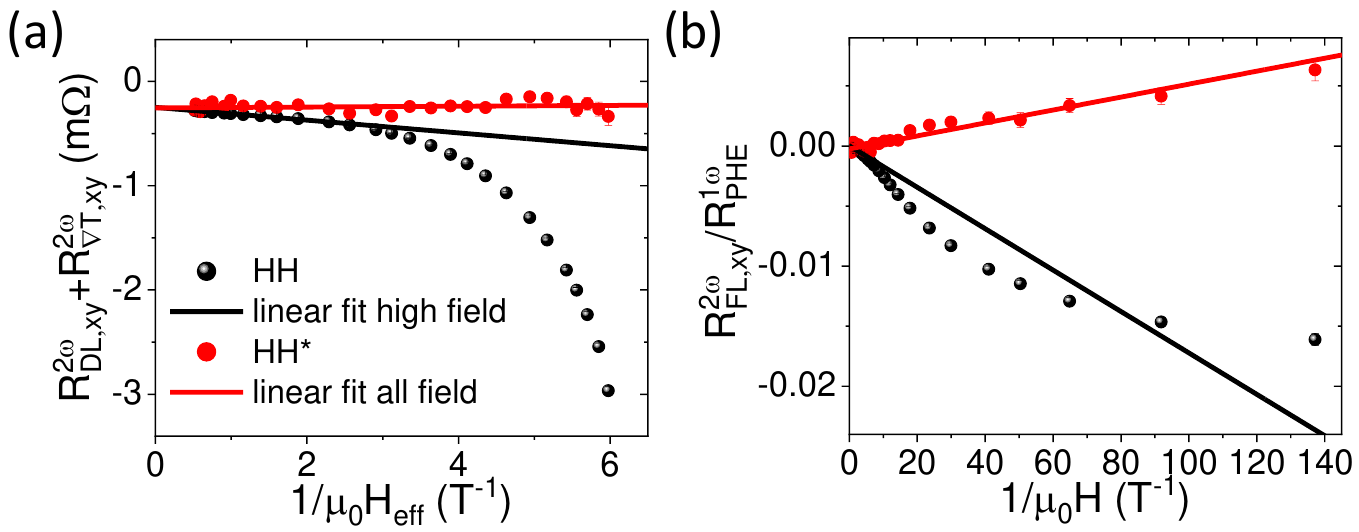}
\caption{\label{fig:figure12} Harmonic Hall resistance analysis in YIG(6.2~nm)/Pt (3~nm) at a current of 4~mA. Dependence of the (a) Transverse DL resistance  with magnetothermal contribution for fields down to 7~mT obtained without and with corrections using Eq.~\ref{eq:fourteen}. (b) Transverse FL resistance divided by the PHE resistance obtained without and with corrections using Eq.~\ref{eq:fifteen}. The solid lines are linear fits to the data, for the uncorrected data the fit is limited to the high field region above 250 mT.}
\end{figure}

Figures~\ref{fig:figure12} (a) and (b) show the variation of $R_{\mathrm{DL}, xy}^{2\omega} + R_{\mathrm{\nabla T}, xy}^{2\omega} $ and $\frac{R_{\mathrm{FL}, xy}^{2\omega}}{R_\mathrm{PHE}^{1\omega}}$ as a function of $(\mu_0 H_{\rm eff})^{-1}$ and $(\mu_0 H)^{-1}$, respectively, at a current of 4~mA. When extracting the DL effective field using $R_{\mathrm{DL}, xy}^{2\omega} = \frac{R_\mathrm{AHE}^{1\omega}}{\mu_{0}H_\mathrm{eff}}B_\mathrm{DL}$, we obtain a positive $B_\mathrm{DL}$ because the sign of the AHE is negative in YIG/Pt. For Pt on top of YIG, a positive sign of $B_\mathrm{DL}$ is unexpected given the positive sign of the spin Hall effect in Pt. The magnitude of $B_\mathrm{DL}$ is also unphysically large, $B_\mathrm{DL}$ = 72 $\pm$ 4~mT at 4~mA. This would be equivalent to a giant and negative spin Hall angle of $-118 \pm 7 \%$, which is of opposite sign and more than one order of magnitude larger than obtained in Pt/CoFeB and Pt/Co. 
Additionally, the FL torque estimated using the standard analysis gives $B_\mathrm{FL}$ = -0.69 $\pm$ 0.03~mT. $B_\mathrm{FL}$ has opposite sign relative to the Oersted field and is considerably larger than previously measured in YIG/Pt using other techniques for a comparable current density~\cite{hahn13, fang17, yang20}. \\

These striking inconsistencies can be understood by taking into account the strong nonlinearities shown in Fig.~\ref{fig:figure12}. According to Eqs.~(\ref{eq:two}) and (\ref{eq:seven}) for the in-plane angular dependence, the conventional harmonic Hall resistance analysis misestimates $R_{\mathrm{DL}, xy}^{2\omega}$ by $R_\mathrm{PHE}^{2\omega}$ and $R_{\mathrm{FL}, xy}^{2\omega}$ by $R_\mathrm{PHE}^{2\omega}/2$. As shown in our joint paper~\cite{jointpaper} the \emph{m}$\!^{\dagger}$\!\emph{m}PHE dominates the second harmonic Hall signal in YIG/Pt. Due to the small $R_\mathrm{AHE}^{1\omega}$ of YIG/Pt, any misestimation of the $R_{\mathrm{DL}, xy}^{2\omega}$ resistance due to the \emph{m}$\!^{\dagger}$\!\emph{m}PHE leads to a giant misestimation of the DL torque, explaining the unphysical values of the DL torque obtained in the high-field limit using the standard harmonic Hall resistance analysis. \\

Using the same method as described previously for conductive FM, we can correct for the \emph{m}$\!^{\dagger}$\!\emph{m}PHE contribution. Such a correction strongly affects the estimation of $R_{\mathrm{DL}, xy}^{2\omega} + R_{\mathrm{\nabla T}, xy}^{2\omega}$ and $\frac{R_{\mathrm{FL}, xy}^{2\omega}}{R_\mathrm{PHE}^{1\omega}}$, as evidenced by the linear fits reported in Fig.~\ref{fig:figure12} (a) and (b). After correction, the variation of $R_{\mathrm{DL}, xy}^{2\omega}$ as a function of field is close to zero and the estimated $B_\mathrm{DL}$ is of -5 $\pm$ 6~mT. Due to the nearly zero variation of $R_{\mathrm{DL}, xy}^{2\omega}$ associated with the small AHE in YIG/Pt it is difficult to obtain a good estimation of the DL torque efficiency. Unfortunately, we could not use MOKE to obtain a reliable estimate of the DL torque due to the small magnetooptical response of YIG. The estimation of the total FL torque gives $B_\mathrm{FL}= 0.22 \pm 0.02$~mT. This value is in excellent agreeement with the estimation in our joint paper using a different method ($B_\mathrm{FL}$ = 0.218 $\pm$ 0.004~mT) \cite{jointpaper} and very similar to the Oersted field. These results evidence the reliability of our approach to estimate the torque. 

\subsection{SOT efficiency and spin Hall conductivity of P\lowercase{t} and W} \label{sec:SHC}

The effective fields $B_{\mathrm{DL}}$ and $B_{\mathrm{FL}}$ measured in the different samples can be compared to other results in the literature by converting them into the SOT efficiencies~\cite{nguyen16,manchon19}
\begin{eqnarray}
    \xi_{\rm DL,FL}^{j}=\frac{2e}{\hbar}{M_\mathrm{s}}{t_\mathrm{F}}\frac{B_{\rm DL,FL}}{J_\mathrm{NM}}
\label{eq:sixteen},\\
       \xi_{\rm DL,FL}^{E}=\frac{2e}{\hbar}{M_\mathrm{s}}{t_\mathrm{F}}\frac{B_{\rm DL,FL}}{E}
\label{eq:seventeen}
\end{eqnarray}
where $e$ is the elementary charge, $\hbar$ the reduced Planck constant, $t_\mathrm{F}$ the thickness of the FM, $M_\mathrm{s}$ the saturation magnetization, $J_\mathrm{NM}$ the current density in the NM layer, and $E$ the applied electric field. The saturation magnetization was measured by SQUID magnetometry on unpatterned samples deposited in the same conditions as the Hall bar samples. The efficiency normalized by the current density, $\xi_{\rm DL,FL}^{j}$, is the ratio of the spin current absorbed by the FM and the injected charge current in the NM. This efficiency corresponds to the effective spin Hall angle of the NM. The efficiency normalized by the applied electric field, $\xi_{\rm DL,FL}^{E}$, is independent of the current distribution in the FM/NM bilayer and corresponds to the effective spin Hall conductivity of the NM.

We compare the DL torque efficiency obtained using the standard harmonic Hall resistance analysis, $\xi_{\rm DL, HH}^{j,E}$, with the one corrected for the \emph{m}$\!^{\dagger}$\!\emph{m}PHE, $\xi_{\rm DL, HH^{*}}^{j,E}$, and measured by MOKE, $\xi_{\rm DL, MOKE}^{j,E}$. For the FL torque we estimated the torque efficiency after subtraction of the Oersted field contribution using the standard harmonic Hall resistance analysis, $\xi_{\rm FL}^{j,E}$ and accounting for the \emph{m}$\!^{\dagger}$\!\emph{m}PHE correction, $\xi_{\rm FL, HH^{*}}^{j,E}$. Table \ref{table1} summarizes the different values of the efficiency for all the samples investigated in this study together with the respective resistance and $M_\mathrm{s}$. Note that the positive sign of the FL efficiency corresponds to an effective field opposite to the Oersted field in our sign convention.\\

The torque efficiency estimated using the harmonic Hall resistance method without correction is larger for CoFeB than for Co. When accounting for the \emph{m}$\!^{\dagger}$\!\emph{m}PHE contribution, the corrected DL torque efficiencies are smaller in all the studied samples and less sensitive to the type of FM. We find $\xi_{\rm DL, HH^{*}}^{j} = 8.3 \pm 0.3$ \% in Pt/CoFeB and 7.4 $\pm$ 0.5 \% in Pt/Co, in excellent agreement with $\xi_{\rm DL, MOKE}^{j}$. In YIG/Pt the DL torque efficiency is estimated to be of 8  $\pm$ 9~\% after correction covering the range of torque efficiencies usually obtained with Pt For the W-based samples, the DL torque efficiency decreases from -42 $\pm$ 6~\% in W/CoFeB and -33 $\pm$ 2 \% in W/Co to -24 $\pm$ 2 \% and -21 $\pm$ 3 \%, respectively, after correction, in much better agreement with the MOKE results.  Using Eq.~(\ref{eq:seventeen}), we obtain spin Hall conductivities of about $3 \times 10^{5}$~$\Omega$ m$^{-1}$ for Pt and $1.8\times 10^{5}$~$\Omega$ m$^{-1}$ for W, with excellent agreement between the corrected harmonic Hall resistance method and MOKE. These values are also in excellent agreement with recent calculations for both Pt~\cite{wang16} and W~\cite{mchugh20, ghosh22}. \\

Using the harmonic Hall resistance method without correction for the FL torque leads to an even larger relative difference of efficiency when comparing the samples with Co and CoFeB. $\xi_{\rm FL, HH}^{j}$ is larger for Pt/CoFeB than for Pt/Co but smaller in W/CoFeB than in W/Co. Strikingly, $\xi_{\rm FL, HH}^{j}= 0.4 \pm 0.3$ \% in W/CoFeB and ten times larger in W/Co (4.3 $\pm$ 0.1 \%). When accounting for the \emph{m}$\!^{\dagger}$\!\emph{m}PHE contribution, the corrected FL torque efficiency is independent of the FM, with $\xi_{\rm FL, HH^{*}}^{j} = 1.5$ \% in Pt and 6.0 \% in W.\\

This extreme sensitivity of the estimation of the torque efficiency  on the magnonic contribution for both Pt and W could explain, at least partially, the variability and inconsistencies in the reported SOT efficiencies ~\cite{manchon19, lau17, karimeddiny20, zhu20, zhu21, karimeddiny23}\\

\begin{table*}

\resizebox{\textwidth}{!}{
\begin{tabular}{l|c|c|c|c|c|c|c|c|c|c|c|c}
\hline\hline
Sample &$R$ &$M_\mathrm{s}$    &$\xi_{\rm DL, HH}^{\textit{$j$}}$&$\xi_{\rm DL, HH^{*}}^{\textit{$j$}}$&$\xi_{\rm DL, MOKE}^{\textit{$j$}}$  
&$\xi_{\rm FL}^{j}$& $\xi_{\rm FL, HH^{*}}^{j}$ &$\xi_{\rm DL, HH}^{E}$ &$\xi_{\rm DL, HH^{*}}^{E}$ &$\xi_{\rm DL, MOKE}^{E}$  
&$\xi_{\rm FL}^{E}$ & $\xi_{\rm FL, HH^{*}}^{E}$  \\
 
 &[$\Omega$] &[kA/m] &[$\%$]& [$\%$]&[$\%$] &[$\%$] &[$\%$] &$10^{5} (\Omega m)^{-1}$ &$10^{5} (\Omega m)^{-1}$ &$10^{5} (\Omega m)^{-1}$ &$10^{5} (\Omega m)^{-1}$ &$10^{5} (\Omega m)^{-1}$\\
\hline\hline

Pt(5)/CoFeB(2.5) & 501  &939 $\pm$ 6 &9.8 $\pm$ 0.2 &8.3 $\pm$ 0.3 & 7.7 $\pm$ 0.3 & 3.0 $\pm$ 0.2 &1.7 $\pm$ 0.1 &3.72 $\pm$ 0.06 &3.1 $\pm$ 0.1 &2.9 $\pm$ 0.1 &1.13 $\pm$ 0.06 &0.65 $\pm$ 0.01 \\
\hline
Pt(5)/Co(2.5) & 552 &1082 $\pm$
10 &8.9 $\pm$ 0.2 &7.4 $\pm$ 0.5 &7.9 $\pm$ 0.1      & 2.2 $\pm$ 0.1  & 1.4 $\pm$ 0.1 & 3.20 $\pm$ 0.07 & 2.7 $\pm$ 0.2 & 2.82 $\pm$ 0.05 & 0.79 $\pm$ 0.02 & 0.50 $\pm$ 0.01 \\
\hline
W(5)/CoFeB(2.5) & 2405 &876 $\pm$
6 &-42 $\pm$ 6 &-24 $\pm$ 2 &-25$\pm$ 2 &0.4$\pm$ 0.3& 6.0$\pm$ 0.1&  -3.0 $\pm$ 0.4 & -1.7 $\pm$ 0.2 & -1.7 $\pm$ 0.10 & 0.03 $\pm$ 0.02 & 0.42 $\pm$ 0.01 \\
\hline
W(5)/Co(2.5)  & 2076 &954 $\pm$ 4 &
-33 $\pm$ 2 &-21 $\pm$ 3  &-20$\pm$ 1 &4.3$\pm$ 0.1 &6.0$\pm$ 0.1 &  -3 $\pm$ 0.2 & -1.9 $\pm$ 0.2 & -1.8 $\pm$ 0.1 & 0.39 $\pm$ 0.01 & 0.55 $\pm$ 0.01 \\
\hline
YIG(6.2)/Pt(3) & 924 &116 $\pm$
12 &-118 $\pm$ 7 &8 $\pm$ 9 & -  & 1.5 $\pm$ 0.1  & 0.1 $\pm$ 0.1 &  -21 $\pm$ 1 & 1 $\pm$ 2 & - & 0.27 $\pm$ 0.02 & 0.02 $\pm$ 0.02\\
\hline\hline
\end{tabular}
}

\caption{\label{table1}  Summary of the resistance, $M_\mathrm{s}$ and SOT efficiencies in the Pt/CoFeB, Pt/Co, W/CoFeB, W/Co and YIG/Pt samples. The four point resistance $R$ is measured between two contacts 100 $\mu$m apart (except for YIG/Pt, 50 $\mu$m apart). The uncertainty on the estimated torque efficiencies represents the standard deviation of the data measured at different currents.} 
\end{table*}
\color{black}

Finally, we can apply the previous knowledge to also quantify the change of the magnetization $\Delta M(I)$ due to current-induced magnon creation-annihilation. To that end, we use the known value of $C_\mathrm{mag}$ to extract $R_\mathrm{PHE}^{2\omega}$ using Eq.~(\ref{eq:thirteen}). As we show in our joint paper, $\frac{\Delta M(I)}{M_\mathrm{s}} = \frac{R_\mathrm{PHE}^{2\omega} }{ 2R_\mathrm{PHE}^{1\omega}}$~\cite{jointpaper}. Using our estimate of $R_\mathrm{PHE}^{2\omega}$ and our measurement of $R_\mathrm{PHE}^{1\omega}$, we evaluate the relative change of the total magnetization at different fields and currents. In Pt/CoFeB and Pt/Co, $\frac{\Delta M(I)}{M_\mathrm{s}}$ remains much below 1\% at the highest current density used in this study for a magnetic field of 20~mT. In particular, we find a change of magnetization of 0.15\% at a current of $7.5$~mA in Pt/CoFeB ($J_\mathrm{NM} \approx 1.4 \times 10^{11}$~A/m$^{2}$), and of 0.09\% at the same current in Pt/Co. The largest change of 0.73\% was obtained in W/CoFeB at 20 mT and a lower current of 6~mA ($J_\mathrm{NM} \approx 1.02 \times 10^{11}$~A/m$^{2}$). The largest change of the magnetization is expected using W as the NM due to its larger spin Hall angle. We note that in Co and CoFeB the relative change of magnetization is much smaller than the one obtained in YIG/Pt, which is of few percents at similar fields and current densities~\cite{jointpaper}. 
 
Our results evidence that even a small change of the total magnetization of less than a percent can lead to a strong misestimation of the spin-orbit torques. This is not unexpected as the change of the magnetization direction due to the spin orbit torques are also small.
\color{black}

 \section{Mitigation of magnonic contributions in the harmonic Hall resistance method} \label{sec:mitigation} 

In this section, we outline when a large magnonic contribution and strong misestimation of the torques areexpected using the standard harmonic Hall resistance technique. We also provide some recommendations to limit torque misestimations when using this method. As we show in Figs.~\ref{fig:figure3}, \ref{fig:figure4}, \ref{fig:figure10} and \ref{fig:figure12} for in-plane magnetized samples, a strong deviation of the second harmonic Hall resistances from the linear dependence on the inverse magnetic field is a clear signature of the \emph{m}$\!^{\dagger}$\!\emph{m}PHE contribution and consequent torque misestimations. Using a narrow range of fields for the measurement and fitting should therefore be avoided. Fitting the second harmonic Hall resistance only in the high field limit is also not sufficient for a proper estimation of the torque in presence of a large magnonic contribution. In any case it is desirable to control for possible magnonic contributions in a wide range of fields. If a nonlinearity is observed, the correction procedure outlined in Sect.~\ref{sec:torques_1} can remove the detrimental influence of the \emph{m}$\!^{\dagger}$\!\emph{m}MRs and provide an accurate estimate of the torques.\\

To avoid large magnonic contributions the FM should have a large damping, a large $M_\mathrm{s}$ and a Curie temperature $T_\mathrm{c}$ much higher than the measurement temperature, usually resulting in a large magnon stiffness. The properties of a FM further depend sensitively on its thickness. Ultra-thin FM layers typically have a reduced $M_\mathrm{s}$ and $T_\mathrm{c}$ compared to the bulk materials~\cite{zhang01}. In particular, sub-nm thick Co and CoFeB can have a $T_\mathrm{c}$ much below their bulk Curie temperature~\cite{schneider90,huang93,chiba11, lee17, zhou20} and thus a large magnon population at room temperature. Using ultra-thin ferromagnets could thus lead to larger contributions of the magnon creation-annihilation.

Moreover, the \emph{m}$\!^{\dagger}$\!\emph{m}PHE contribution is not only proportional to the relative change of the magnon population but also to the change of transverse resistance due to the PHE, $R_\mathrm{PHE}^{1\omega}$. A magnetic layer with a PHE much smaller than the AHE, such as a rare earth ferrimagnetic alloy~\cite{zhou18} would be ideal to limit the \emph{m}$\!^{\dagger}$\!\emph{m}PHE contribution. On the other hand, these systems often show self-torques ~\cite{cespedes-berrocal21}, which make it hard to evaluate all the SOT contributions. \\

Following our discussion above, we provide guidelines to limit the errors on the estimations of the SOTs in future studies relying on the harmonic Hall resistance technique: 

\begin{itemize}
\item Use FMs with large damping and $M_\mathrm{s}$ at the measurement temperature to limit the magnon creation-annihilation. 

\item Measure at a temperature far below $T_\mathrm{c}$. Ideally perform measurements as a function of temperature.

\item Use FMs with a large AHE compared to their PHE. Materials with large PHE compared to their AHE such as permalloy (Ni$_{80}$Fe$_{20}$)~\cite{mcguire75}, or a large SMR compared to their AHE, such as YIG/Pt, should be avoided for reliable DL torque measurements. 

\item Use a range of fields as broad as possible to inspect the field dependence of the DL and FL contributions, from few dozens of mT, large enough to avoid contribution from magnetic anisotropies, to at least 1~T, a field large enough to confirm the field dependence.

\item Measure both the longitudinal and transverse second harmonic resistances and correct the estimated torque as outlined in Sect.~\ref{sec:torques_1} if nonlinearities are observed.
\color{black}
\end{itemize}

There is no perfect FM for torque measurements using the harmonic Hall resistance method without accounting for the \emph{m}$\!^{\dagger}$\!\emph{m}PHE, as no FM is fully immune to current-induced magnon creation-annihilation processes or has a vanishing PHE. A few nm-thick Co layer with high $T_\mathrm{c}$, $M_\mathrm{s}$, damping and a large AHE appears as an optimal solution when accounting for all the above recommendations, but remains an imperfect choice. As we showed in Sect.~\ref{sec:torques_2} in the case of W/Co, the DL torque in $2.5$-nm-thick Co is still overestimated by 45\% at room temperature.\\

In this work, we only considered in-plane magnetized samples using angular-dependent Hall resistance measurements ~\cite{avci14}. However, the magnonic contribution is not limited to this measurement configuration. As long as the magnetization has a component along the spin accumulation direction, it can give rise to nonzero nonlinear effects associated with magnon creation-annihilation. This can affect the estimation of the torque for FM with perpendicular magnetic anisotropy (PMA). The usual experimental configuration for PMA magnets is to measure the first and second harmonic Hall resistances when applying a magnetic field along the $x$ and $y$ direction and consider contributions due to the DL torque, FL torque and magnetothermal effects. This type of measurement is mostly performed at low external field where the magnetization is nearly out of plane ~\cite{garello13,hayashi14, ghosh17, krishnia21}. Due to the out-of-plane component of the magnetization, the nonlinear anomalous Hall effect, namely the \emph{m}$\!^{\dagger}$\!\emph{m}AHE, can also contribute when the field is applied along the $y$ direction. This contribution would also have an angular and field dependence similar to torque contributions when the field is applied along $y$, possibly leading to the misestimation of both the DL and FL torques. Recently, a discrepancy between optical and electrical detection of torques in thin PMA magnet was evidenced, but the artifacts were of unidentified origin~\cite{karimeddiny23}. A large \emph{m}$\!^{\dagger}$\!\emph{m}AHE contribution could be the source of this discrepancy.

In general, the spin accumulation can be along $y$ but also along $x$ and $z$, as demonstrated in materials with low symmetry ~\cite{macneill17,liu21b,roy22}. These unconventional spin accumulation will lead to additional contributions to the \emph{m}$\!^{\dagger}$\!\emph{m}MRs that will affect the estimation of the SOTs using the harmonic Hall resistance method. These contributions are discussed in our joint paper~\cite{jointpaper}. \\

\color{black}
\section{Conclusions} \label{sec:end}

In summary, we evidenced that nonlinear effects associated with current-induced magnon creation-annihilation magnetoresistances appear prominently in the second harmonic longitudinal and transverse resistances of different NM/FM bilayers. The \emph{m}$\!^{\dagger}$\!\emph{m}PHE leads to a strong misestimation of the spin-orbit torques in NM/FM bilayers with in-plane magnetization measured by the harmonic Hall resistance method. This misestimation is particularly significant in FM layers with small damping, $M_\mathrm{s}$ and $T_\mathrm{c}$, as well as in FM with a relatively large PHE compared to the AHE. We found that the DL torque is overstimated by 15\% in Pt/Co, 30\% in Pt/CoFeB, 100\% in W/CoFeB, and more than 1000\% with the wrong sign in YIG/Pt if the \emph{m}$\!^{\dagger}$\!\emph{m}PHE is not properly accounted for. \\

We proposed a revised analysis of the harmonic Hall resistances in order to account for the magnonic contribution. The values of the DL and FL torque and corresponding efficiencies corrected for the \emph{m}$\!^{\dagger}$\!\emph{m}PHE are consistent in sign and magnitude across different FM and NM layers, supporting the validity of our analysis. Furthermore, the DL torques measured using the corrected harmonic Hall resistance method are in excellent agreement (within 5\%) with MOKE measurements performed on the same samples and calibrated by NV magnetometry. A comprehensive comparison of the DL and FL torque measured using these different methods is presented in Table~\ref{table1}. \\

Our results evidence that accounting for the \emph{m}$\!^{\dagger}$\!\emph{m}PHE is essential for the proper evaluation of the SOTs, particularly for FM with "soft" magnetic properties and/or weak AHE. The \emph{m}$\!^{\dagger}$\!\emph{m}PHE, which depends markedly on the composition and thickness of FM/NM bilayers, can likely explain several discrepancies observed in the literature regarding the amplitude and sign of the SOTs.  The considerations and methodology presented here naturally extend to the measurement of orbital torques. \color{black}\\

As current-induced magnon creation-annihilation effects are ubiquitous in NM/FM systems, future work could address the influence of the \emph{m}$\!^{\dagger}$\!\emph{m}MRs on other techniques used to measure the SOTs, in particular those where magnetoresistive effects and the assumption of constant magnetization play a role. Accounting for all measurement contributions not due to torques is an essential step to obtain a complete understanding of the physical mechanisms underlying the SOTs and key for the optimization of SOT devices. 

\begin{acknowledgments}
We acknowledge Morgan Trassin for providing the YIG/Pt sample. Discussions with Sa\"ul V\'elez are gratefully acknowledged. This work was supported by the Swiss National Science Foundation (Grant No. 200020\_200465). P.N. acknowledges the support of the ETH Zurich Postdoctoral Fellowship Program 19-2 FEL-61.
\end{acknowledgments}

\nocite{*}
\bibliographystyle{unsrt}
\bibliography{apssamp}

\end{document}